\newcommand{\MarkChange}{
}

\documentclass[acmsmall]{acmart}

\AtBeginDocument{%
 }

\setcopyright{rightsretained}
\acmJournal{PACMHCI}
\acmYear{2023} \acmVolume{7} \acmNumber{CSCW1} \acmArticle{134} \acmMonth{4} \acmPrice{}\acmDOI{10.1145/3579610}

\makeatletter
\gdef\@copyrightpermission{
  \begin{minipage}{0.2\columnwidth}
   \href{https://creativecommons.org/licenses/by/4.0/}{
   \includegraphics[width=0.90\textwidth]{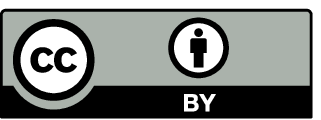}}
  \end{minipage}\hfill
  \begin{minipage}{0.8\columnwidth}
   \href{https://creativecommons.org/licenses/by/4.0/}{This work is licensed under a Creative Commons Attribution International 4.0 License.}
  \end{minipage}
  \vspace{5pt}
}
\makeatother

\usepackage{amsmath}
\usepackage{amsthm}
\usepackage{bbold} %
\usepackage{mathtools}
\usepackage{color}
\usepackage[normalem]{ulem}

\theoremstyle{definition}
\newtheorem{definition}{Definition}[section]
\theoremstyle{definition}

\newtheorem{lemma}{Lemma}[section]
\newtheorem{theorem}{Theorem}

\title[Platform-supported Auditing of Social Media Algorithms for Public Interest]{Having your Privacy Cake and Eating it Too: Platform-supported Auditing of Social Media Algorithms for Public Interest}
\author{Basileal Imana}
\affiliation{%
  \institution{University of Southern California}
  \city{Los Angeles}
  \state{CA}
  \postcode{43017-6221}\country{USA}
}
\author{Aleksandra Korolova}
\affiliation{%
  \institution{Princeton University}
  \city{Princeton}
  \state{NJ}
  \postcode{08540}\country{USA}
}
\author{John Heidemann}
\affiliation{%
  \institution{USC/Information Science Institute}
  \city{Los Angeles}
  \state{CA}
  \postcode{43017-6221}\country{USA}
}



\allowdisplaybreaks %

\def\Snospace~{\S{}}

\makeatletter
  \newcommand\EatSpacesHack{\@bsphack\@esphack}
\makeatother
  \renewcommand{\comment}[1]{\EatSpacesHack}
  \newcommand{\PostSubmission}[1]{\EatSpacesHack}
  \newcommand{\todo}[1]{\EatSpacesHack}
  \newcommand{\basi}[1]{\EatSpacesHack}
  \newcommand{\ak}[1]{\EatSpacesHack}
  \newcommand{\reviewfix}[1]{\EatSpacesHack} 

\begin{document}

\begin{abstract}
\reviewfix{}
Social media platforms 
  curate access to information and opportunities,
  and so play
  a critical role in shaping public discourse today.
The opaque nature of the algorithms
  these platforms use to curate content
  raises societal questions.
Prior studies have
  used black-box methods led by experts or collaborative audits
  driven by everyday users
  to show that these algorithms
  can lead to biased or discriminatory outcomes.
However, existing auditing methods
  face fundamental limitations because they 
  function independent of the platforms.
Concerns of potential harmful outcomes have
  prompted proposal of legislation in both the U.S.~and the E.U.~to
  mandate a new form of auditing where vetted external researchers
  get privileged access to social media platforms.
Unfortunately, to date there have been no concrete technical proposals
  to provide such auditing,
  because auditing at scale risks disclosure of
  users' private data and platforms' proprietary algorithms.
We propose a new method for \emph{platform-supported auditing}
  that can meet the goals of the proposed legislation.
The first contribution of our work
  is to enumerate the challenges and the limitations of existing auditing methods
  to implement these policies at scale.
Second, we suggest that limited, privileged access to
  \emph{relevance estimators}
  is the key to enabling generalizable
  platform-supported auditing of social media platforms
   by external researchers.
Third, we show platform-supported auditing need not risk
  user privacy nor disclosure of platforms' business interests
  by proposing an auditing framework that protects against
  these risks.
For a particular fairness metric, we show that ensuring privacy
  imposes only a small constant factor increase
  ($6.34\times$ as an upper bound, and $4\times$ for typical parameters)
  in the number of samples required for accurate auditing.
Our technical contributions,
 combined with ongoing legal and policy efforts,
 can enable public oversight into how social media platforms affect individuals and society
 by moving past the privacy-vs-transparency hurdle.
\end{abstract}

\maketitle

\section{Introduction}
  \label{sec:intro}

{ \MarkChange

\reviewfix{}
Social media platforms are no longer just
  digital tools that connect friends and family---today they play a critical role in   
  shaping public discourse and moderating access to information
  and opportunities.
Platforms such as
  Facebook, Instagram, Twitter, LinkedIn and TikTok
  have become the new search engines~\cite{TikTokSearch, TikTokSearch2},
  helping individuals find important life opportunities such as jobs~\cite{Hosain2021, Nikolaou2014},
  and are often sources of news and advice~\cite{Shearer2017}.
Content curation on these platforms is done by
  \emph{algorithms that estimate relevance of content to users}, which 
  raises fundamental questions about their societal
  implications.
\reviewfix{}
For example:
Do these algorithms discriminate against certain demographic
  groups?
Do they amplify political content in a way that threatens democracy?
Does optimization for relevance, often defined via engagement, emphasize extreme viewpoints,
  dividing society?

Prior studies have used either user-driven or expert-led
 audits to answer some of these questions but their methodologies
 have limitations.
User-driven, collaborative approaches rely
  on day-to-day observations by users
  to infer algorithmic
  behavior
  of the platforms~\cite{Eslami2019, Shen2021, DeVos2022, Birhane2022}.
Such audits have uncovered important societal issues
  such as racial bias in image cropping  on Twitter~\cite{TwitterImageCrop, Birhane2022}.
Although user-driven audits can detect   
  such visible algorithmic results,
  other types of harm can be invisible to end-users
  and require systematic study by experts to detect~\cite{DeVos2022}.
  
Researchers and journalists
  have performed audits in a more systematic way to uncover 
  harm on platforms.
Examples of their findings include biased or
  discriminatory ad targeting and delivery~\cite{Lecuyer2015, Ali2019a,Asplund2020, Imana2021},
  amplification of hateful content~\cite{WsjFBFilesHateful2021},
  political polarization~\cite{Ribeiro2020, WsjFBPolarization2021, TwitterAlgAmplification, Haroon2022, papakyriakopoulos2022algorithms},
  and promotion of addictive behavior in teens~\cite{WsjFBFilesTeens2021}
\reviewfix{}.
We describe harms in detail in~\autoref{sec:relevance_harms}.
Although existing methods that auditors use have been crucial in
  uncovering harms and driving change,
  they are reaching hard limits
  in terms of what they can reliably and provably learn about the
  role of platforms' algorithms~\cite{Ali2019a, Ali2019b, Imana2021}.
We expand on limitations of both user- and expert-driven methods in \autoref{sec:improve_limitations}.
}

Concerns of algorithmic harms
  and challenges of existing auditing methods
   have prompted policy proposals for a new type of auditing~\cite{Pata2022, Sdaa2021, Aaa2019, Ajopta2021, Aia2021, dsa2021}.
Such legislation is motivated by the critical role of social
  media today because of personalization algorithms,
  the need for auditing these algorithms to mitigate harms,
  and the concern that self-auditing has not addressed these issues.
\reviewfix{}
{\MarkChange
In addition to legislation,
  the White House recently reported
  on civil rights of users in the digital age,
  raising similar challenges~\cite{WhiteHouseBillofRights}.
}
Legislative proposals have suggested platforms
  make data available to vetted, external researchers,
  who will audit platforms' algorithms and evaluate
  their alignment with societal or legal
  expectations.
We call such proposals
  \emph{platform-supported auditing},
  and discuss how they extend existing auditing taxonomies 
  in~\autoref{sec:overview_framework}.
Two such prominent proposals are
  U.S.'s~Platform Accountability and Transparency Act (PATA~\cite{Pata2022})
  and the E.U.'s~Digital Services Act (DSA~\cite{dsa2021}).
The E.U.~will start enforcing DSA in 2024~\cite{DsaTimeline},
  and PATA is still a proposed bill.
DSA will require platform to support auditing
  but how the proposal can be implemented in practice
  remains an open question~\cite{Nonnecke2022}.

A critical concern outlined in the legislative proposals is the need to protect
  \emph{privacy}.
Privacy encompasses protecting both users' profile data
  and platforms' proprietary algorithms.
  Prior research has shown increasing transparency
  while protecting users' data and platforms' proprietary algorithms
  presents technical challenges~\cite{Bogen2020, metatech, austin2021race, TwitterAlgAmplification, Nonnecke2022, Andrus2021}.
Platforms such as Facebook have also cited
  privacy of users' data as a constraint on increasing their transparency
  efforts~\cite{FacebookBlockedNYU, Vermeulen2021}.
In our discussions with platforms, privacy of their proprietary algorithms is also
  an issue that quickly comes up.
Until now, no actionable proposals
  have been put forth for how to implement platform-supported
  auditing while addressing the privacy concerns.
The challenge is to translate the ambitious policy goals into a practical
  and scalable implementation.

\reviewfix{}
{\MarkChange
Our first contribution is to \emph{enumerate limitations of existing
  auditing methods for implementing
  platform-supported auditing at scale} (\autoref{sec:problem_statement}).
}
We start with an overview of what DSA and PATA
  compel social media platforms to make available to external auditors.
We then enumerate the significant limitations and
  non-generalizability of existing external auditing methods to study
  algorithmic harms on these platforms.
Specifically, although existing methods have been crucial to
  detecting how various social media platforms harm different demographic
  groups and our society at large, they
  do not generalize well to study multiple
  types of harms, demographic groups or platforms.

{\MarkChange
Our second contribution is to suggest that
  \emph{transparency of relevance estimators is the key
  to enabling a generalizable and actionable framework
  for platform-supported auditing} (\autoref{sec:why_relevance}).
  }
Our proposal provides a plausible, practical approach to
  platform-supported auditing.
While the DSA and PATA require auditing legislatively,
  they do not specify a mechanism;
  our approach is the first to meet this need.
We show the importance of auditing relevance estimators
  by examining platforms' documentations that show
  they are the ``brains'' that shape
  delivery of every piece of  organic and promoted content
  on social media.
Despite being the core drivers,
  these algorithms are used across multiple social-media platforms
  with little transparency 
  into their definitions of relevance or the specific inputs they use to optimize for it.
We survey prior audits that indirectly measured
  how the use of relevance estimators can result in harmful outcomes
  to show a means to directly query and audit these
  algorithms is the key to increasing transparency and providing a meaningful path to verifying alignment with societal and legal expectations.

{\MarkChange
Our third contribution is to show platform-supported auditing
  \emph{need not risk user privacy nor disclosure of platforms' business interests}.
  }
In \autoref{sec:platform_supported},
  we propose an auditing framework that protects against
  these risks.
Our framework uses the rigorous definition of Differential Privacy (DP) to
  protect private information about \emph{audit participants} from leaking to the \emph{auditor}. 
It also protects the platform by not exposing details
  of the ranking algorithm---the platform shares with the auditor only
  the privatized scores of the relevance estimator, not proprietary source code, models, training data
  or weights.
At a high level (\autoref{fig:auditing_overview}), our framework works as follows:
  an auditor queries the algorithm with a trial content and a list of users
  whose sensitive demographic attributes are known to the auditor.
The platform then calculates how relevant the content is to each user,
  applies a differentially private mechanism to protect information
  that the relevance scores may leak about the users,
  and returns a distribution of noisy scores to the auditor.
Finally, the auditor uses the noisy scores
  and an applicable fairness metric
  to test for disparity
  between the distributions of relevance scores
  the algorithm assigns to different demographic groups.
The auditor chooses the specific type of content, attribute of users and metric of
  fairness to use
  depending on the specific scenario and the type of bias or
  harm they are studying.

We show that the privacy guarantees in our framework
do not prevent an auditor from achieving the same statistical confidence in their analysis as without privacy protections
-- the ``cost of privacy" is an
  increase in the number of samples required for an audit
  by a small constant factor (\autoref{sec:privacy_tradeoffs}). 
We theoretically analyze the trade-off between
  guaranteeing privacy and the minimum sample size required for auditing
  in one concrete scenario -- measuring bias in the delivery of employment ads. 
For the specific fairness metric we study, equality of opportunity,
  we find that the noise that the platform adds
  to guarantee DP increases
  the required sample size by only approximately a factor of 4
  for reasonable auditing parameters,
  and with a strict upper bound of 6.34.
Our contribution is application of a standard DP algorithm to ad relevance estimators,
  a new application where DP enables viable privacy-utility trade-offs.

Overall, our technical contributions show a path exists from
  the proposed legislation to a realizable auditing system.
While full implementation of our framework is future work
  and will require collaboration with a platform,
  conceptually demonstrating how to enable public oversight while
  protecting privacy is an
  important step forward.
We summarize the limitations of our framework in \autoref{sec:model_limitations},
  but as the first proposed solution for implementing DSA- and PATA-like laws,
  it provides a useful starting point for exploring a new solution space.

\section{The Need for A Generalizable Auditing Framework}
   \label{sec:problem_statement}
 
 We discuss recent developments in policies
   that are pushing to increase transparency of
   digital platforms,
   the need for safeguards to protect privacy,
   and the insufficiency of existing auditing methods
   to practically implement these policies at scale.
 
 \subsection{Policy Pushes to Increase Transparency while Ensuring Privacy}
   \label{sec:policy_push}

As  social media platforms increasingly shape
  economic, social and political discourse,
  new policies are being proposed to regulate them.
We discuss two prominent pieces of legislation to mandate independent oversight
  and transparency research on platforms:
  Platform Accountability and Transparency Act (PATA~\cite{Pata2022, Persily2021}, proposed in the US)
  and Digital Services Act (DSA~\cite{dsa2021}, to be enforced in the EU starting in 2024).

PATA proposes mandating platforms to support independent
  research on algorithmic transparency.
A discussion draft was originally proposed in December 2021~\cite{Pata2021},
  followed by a formal introduction of the bill in December 2022~\cite{Pata2022}.
The proposal covers all large platforms
  with at least 25 million unique monthly users.
It mandates the platforms make
  data available to ``qualified researchers''
  who will study how
  platforms negatively impact individuals and society.
Only researchers and projects vetted
  and approved by the National Science Foundation (NSF)
  will be allowed to access platforms' data.
  
DSA was proposed in the EU in December 2020
  to regulate digital platforms and services~\cite{DsaTimeline}.
DSA covers a broader set
  of entities beyond large social media platforms,
  including online marketplaces and app stores.
While DSA has a broader scope than PATA,
  it similarly mandates platforms to allow scrutiny
  of their algorithms by ``vetted researchers'' (Article 40~\cite{dsa2021}).
DSA was approved and passed as a law in November 2022~\cite{DsaTimeline}.

Both PATA and DSA recognize the need to
  ensure safeguards to protect privacy during platform auditing.
PATA emphasizes user privacy, with the necessity to
  ``establish reasonable privacy and cybersecurity safeguards''
   for user data, and to ensure the data platforms provide is
   ``proportionate to the needs of the [\ldots] researchers to
   complete the qualified research project''~\cite{Pata2022}.
DSA acknowledges platforms' desire for ``protection of confidential information,
  in particular trade secrets''~\cite{dsa2021} when
  conducting audits.
To mitigate the risks to users and platforms,
  both proposals require vetting auditors,
  their projects, and results before
  they are published.
Prior to our work, no actionable technical proposals put forth
  methods to implement
  such auditor
  access while protecting users' privacy
  and platforms' proprietary algorithms.

Platforms themselves also often cite their need to protect user privacy as a handicap for their transparency and self-policing capabilities~\cite{austin2021race, metatech}.
For example, Facebook has argued that laws such as the EU's GDPR
  constrain their efforts to make data available to
  researchers~\cite{Vermeulen2021, SocialScienceOne}.
In line with this argument,
  Facebook has constrained transparency efforts
  through actions such as
  providing data without sufficient granularity and accuracy
  needed to conduct meaningful audits through its Ad Archive APIs~\cite{Edelson2019, Vermeulen2021, papakyriakopoulos2022algorithms},
  and shutting down
  accounts used for transparency research by NYU's Ad
  Observatory project~\cite{FacebookBlockedNYU}.
In a partnership Facebook made with Social Science One,
  Facebook cited GDPR concerns and agreed to share data
 only using a differentially private mechanism~\cite{SocialScienceOne}.
Other social media platforms such as Twitter have also raised
  the challenges of sharing data for auditing for societal benefit while protecting
  the privacy of users~\cite{TwitterAlgAmplification}.

{\MarkChange
\reviewfix{}
Overall, the legislative proposals demonstrate society's need
  for increasing transparency.
}
Given policymakers' and platforms' concern about privacy,
  implementing these proposals requires solving
  methodological challenges to increasing transparency
  while safeguarding the privacy of users.
Our auditing framework (\autoref{sec:platform_supported})
  suggests that these policy requirements can be made concrete and viable
  with our proposed methodology.

\subsection{Existing External Auditing Methods are Insufficient}
  \label{sec:improve_limitations}

Until the present, societal and individual harms of social media algorithms have mostly been merely hypothesized or, in some cases, demonstrated by end-users, journalists and researchers through audits done independent of the platforms.

However, such fully external auditing methods are reaching hard limits
  in terms of what they can reliably and provably learn about the optimization algorithms' role; increasing public-interest researchers' calls for legislation and other transparency sources that can support their efforts~\cite{Ali2019a, Ali2019b, Imana2021, datta2018discrimination, Speicher2018, edelson2021universal, papakyriakopoulos2022algorithms}.
Specifically, the fully external auditing methods face fundamental challenges
  accounting for confounding variables,
  measuring the effect of algorithmic decisions that are opaque to end-users,
  and using proxies for sensitive attributes of interest.
As a result, they are difficult to generalize and
  have high cost.
In addition, they are susceptible to platform interference (\autoref{sec:beginning-to-favor}).
We next expand on these challenges.

\textbf{Confounding variables:}
The first challenge is controlling for
  variables that confound measurements.
These confounding factors are present because
  platforms' algorithms operate in an environment that is influenced
  by actions of both users and the algorithms themselves.
These hidden variables make it difficult to attribute
  measured effects to decisions made by platforms' algorithms.

Auditing for bias in ad delivery provides an illustration of
  the challenge of accounting for confounding factors.  
Several factors may drive differences in ad delivery to individuals from different demographic groups, such as
different levels of market competition from other ads for members of different groups, 
as well as 
  differences in platform's use or interaction patterns among users from various demographics.
An external auditor aiming to isolate the role of the relevance estimator for differences in outcomes between demographic groups must control for such factors. 
Designing auditing methods with such controls in place,
  however, is a laborious process that requires careful reasoning and creative hacks.
In particular, it took many years of research effort to get from
  Sweeney's study that gave the first evidence of
  biased ad delivery in 2013~\cite{Sweeney2013}
  to Ali and Sapiezynski et al.'s study that attributed such bias to
  the role the platform's algorithms play in 2019~\cite{Ali2019a}, and the Imana et al.'s 2021 study~\cite{Imana2021} that established that the algorithms are not merely biased, but, in fact, discriminatory.
  
Similar factors can confound measurements of potential harms in personalized organic content delivery.
For example, a study on Twitter used sock-puppet accounts to compare
  their reverse-chronological and personalized timelines,
  and showed Twitter's algorithms distort
  information that users get exposed to~\cite{Bartley2021}.
However,
  the study identifies the duration sock-puppet accounts stay logged-in for and the timeline scrolling capabilities as potential confounding
  factors that could possibly alter the conclusions~\cite{Bartley2021}.
Even Twitter's internal audit of disparate algorithmic amplification
  of political content, for political right compared to political left,
   shows the limits of current methods~\cite{TwitterAlgAmplification}.
The study showed that their metric of amplification,
  which is based on number of impressions, 
  demonstrates the presence of bias on Twitter,
  but that confounding factors prevent any conclusions about potential sources of this bias.

These examples demonstrate the limits of impression-based
  measurements for isolating algorithmic effects.
To increase transparency
  beyond what we have already learned through existing
  external auditing methods,
  a new level of access is needed for auditors
  (\autoref{sec:why_relevance}).
  
  { \MarkChange
\reviewfix{}
\textbf{Opacity to end-users:}
Another challenge is that the effects of platform algorithms
  are often not obvious or visible to end-users.
Collaborative methods that rely
  on end-users' day-to-day experiences
  may not be able to detects harms
  that are invisible or unnoticeable to users~\cite{DeVos2022, Shen2021}.
We discuss this limitation in more detail in~\autoref{sec:related_work}.
}
 
  \textbf{Reliance on proxies:}
A third challenge is the need for an auditor
  to use proxies for demographic attributes that platforms do not
  collect or report.
Auditors may be interested in studying
  the impact of a specific demographic feature on algorithmic personalization,
  but often conduct external audits
  by posing as a regular user or advertiser.
Operating as a normal user or advertiser
  is relatively easy and allows audits without a platform
  support or knowledge,
  but it also means the auditor can only use
  data points that a platform makes available to any user.
For example, in the context of ad delivery,
  some platforms may not report ad impression rates broken down
  by attributes such as gender, race or political affiliation.
Past audits have worked around this challenge by using
  proxies for demographic attributes that platforms do not report~\cite{Ali2019a, Ali2019b, Imana2021}.
However, such workarounds introduce measurement errors~\cite{Imana2021}
  and significantly limit the ability to vary the attributes.
  
 \textbf{Lack of generalizability:}
Another challenge is that existing external auditing methods are often
  not generalizable beyond
  the limited context which they were originally designed for.
For example,
  we carried out a study aiming to ascertain whether job ad delivery algorithms are discriminatory that built upon Ali and Sapiezynski et al.'s work,
  but 
  adding new controls for job qualifications across genders that required additional knowledge about gender composition of current employees of several companies~\cite{Imana2021}.
The use of additional data on employers and the gender of their employees means this method
  does not directly generalize to auditing for discrimination in ad delivery of other types of ads (for example, housing ads) and 
  along other demographic attributes (such as race).
This lack of generalizability is also directly related to the limitations
  of confounding variables and use of proxies discussed above.
In order to work around these limitations,
  researchers often use one-off hacks that are
  experiment- or platform-specific.
Examples include use of random phone
  numbers to generate a random custom ad audience~\cite{Ali2019a},
  and use of public data sources such as
  voter data to build audiences with a specific
  demographic make up~\cite{Speicher2018, Ali2019a, Ali2019b}.
Such public data sources are extremely limited and subject individuals to participation in experiments without their knowledge.

On the other hand, crowdsourced audits that rely on browser extensions
  do not easily generalize beyond desktop versions of platforms,
  a significant limitation to their applicability given that most people today access social media through their phones.
For example, 98.3\% of Facebook users
  access it using a phone app~\cite{GlobalSocialDataReport}.
Furthermore, such extensions need to be customized
  for each platform, and need to be regularly
  maintained to adapt to changes on platforms' websites.

\textbf{Cost of auditing:}
Finally, existing external auditing methods can also incur high
  costs in terms of both time and money.
For ad delivery,
  the state-of-the-art method for auditing involves
  registering as an advertiser,
  running real ads,
  and measuring how they are delivered in real-time
  while controlling for confounding factors~\cite{Ali2019a, Imana2021}.
The monetary cost for this procedure
  can easily accumulate with repeated
  assessments of a platform
  to confirm results over time,
  increase statistical confidence,
  or vary study parameters.
In addition, controlling for confounding factors
  and proxies for measuring delivery along sensitive attributes
  requires time for study design.

For studies of personalization of organic content,
  creation of sock-puppet accounts is expensive because
  it often requires separate hardware and phone number
  verification, and it takes time and effort to make a
  sock-puppet's account activity ``realistic". 
   
Overall, these challenges motivate our approach:
   by using platform-supported auditing centered on relevance estimators,
   we directly focus on platform choices,
   side-stepping confounding variables and proxies.
Explicit platform support also avoids platform interference and minimizes cost,
  provided platforms collaborate, as we explore next.

\subsection{Platforms Beginning to Favor Platform-supported Audits?}
  \label{sec:beginning-to-favor}

\reviewfix{}
 { \MarkChange
Platform-supported audits, of course, require support from the platform,
  so we next look at evolution of the platforms' responses to requests for auditing.

\textbf{Resistance from platforms:}
Traditionally, a major challenge for external auditing methods has been
  resistance from platforms, often citing privacy concerns or violation of their terms of service.
  
External audits collect data either through interfaces
  the platforms provide or by using tools such as
  customized scrapers and browser extensions.
Regular website changes complicate long-term maintenance of
   automated tools that track platforms~\cite{Bartley2021}.
Facebook has resisted external auditing by explicitly blocking
   accounts used to conduct audits~\cite{FacebookBlockedNYU},
   tweaking its APIs to break auditing tools~\cite{FacebookPropublicaBlocked},
   and threatening legal actions against researchers who scrape
   data from its platform~\cite{FacebookBlockedNYULegal}.

\textbf{A change of heart?}
Recently platforms have begun releasing data or providing APIs to researchers,
  suggesting platforms themselves may be interested in some form of platform-supported auditing.
Platform support allows them to manage auditing,
  and perhaps preempt adversarial black-box audits, lawsuits, and explicit regulation.

Facebook announced,
  in its June 2022
  settlement with the U.S.~Department of Justice,
  that it will work towards de-biasing its algorithms
  used to deliver job, housing and credit ads~\cite{FacebookvsHUD, FacebookvsHUD2},
  and deployed changes
  for housing ads in January 2023~\cite{FacebookvsHUD2023}.
The settlement requires Facebook to work with a vetted, external
  entity to verify the changes implemented to its algorithms
  comply with the non-discrimination goals set by the settlement,
  a compliance structure similar to platform-supported
  methods proposed in PATA and DSA.

Multiple platforms have also recently established programs to provide vetted researchers with
  access to data about their algorithms.
In 2021,
  Facebook announced the ``Facebook Open Research \& Transparency'' (FORT) initiative,
that provides privacy-protected datasets and APIs for researchers, so that ``the public can learn more about Facebook's impact on the world from credible and independent academic sources''~\cite{FacebookFORT}.
In the same year, Twitter provided source code of their image salience algorithm to researchers,  challenging them to evaluate it for bias~\cite{Yee2021image, TwitterBiasBounty}.
And in July 2022, YouTube announced ``YouTube Researcher Program'' (YRP),
  which promises to provide academic researchers with data, tools, and support
  to study YouTube and its impact~\cite{YoutubeResearchProg}.
These steps are promising responses
  partially in the direction of proposed legislative requirements.
They suggest
  platforms are considering
  explicit support of methods that increase transparency of their influence on individuals and society.

However, for both YRP and FORT,
  current data available to researchers
  is limited to \emph{public data corpora},
  such as public videos, pages, posts and comments~\cite{YoutubeResearchProg, FacebookFORT}.
While such access is an important first
  step for helping understand how the platforms shape
  public discourse,
  we argue in~\autoref{sec:why_relevance} that it is also
  important for platforms to provide a means to studying
  how their \emph{algorithms} curate (often personalized)
  delivery of content.
We hope our work encourages platforms to expand
  these first efforts to allow researchers to study
  how their algorithms shape access to content. 
}

\section{Relevance Estimators are the Key to Increasing Transparency}
  \label{sec:why_relevance}

In this section, we show transparency of relevance estimators
  is the key to enabling generalizable
  auditing of platforms for potential harms.
To support this claim,
  we first document the importance of relevance
  estimators for content prioritization.
We then survey studies that have shown harmful outcomes that result
  from use of these algorithms.

\subsection{Relevance Estimators: ``Brains'' of Social Media Platforms}
  \label{sec:relevance_brain}

Relevance estimators are algorithms that form the primary
  means by which platforms select every piece of
  content shown to users.
Prior work and platform documentation show the importance of these algorithms,
  but provide little transparency into how they operate.

\begin{figure}
  \centering
  \includegraphics[width=0.75\columnwidth]{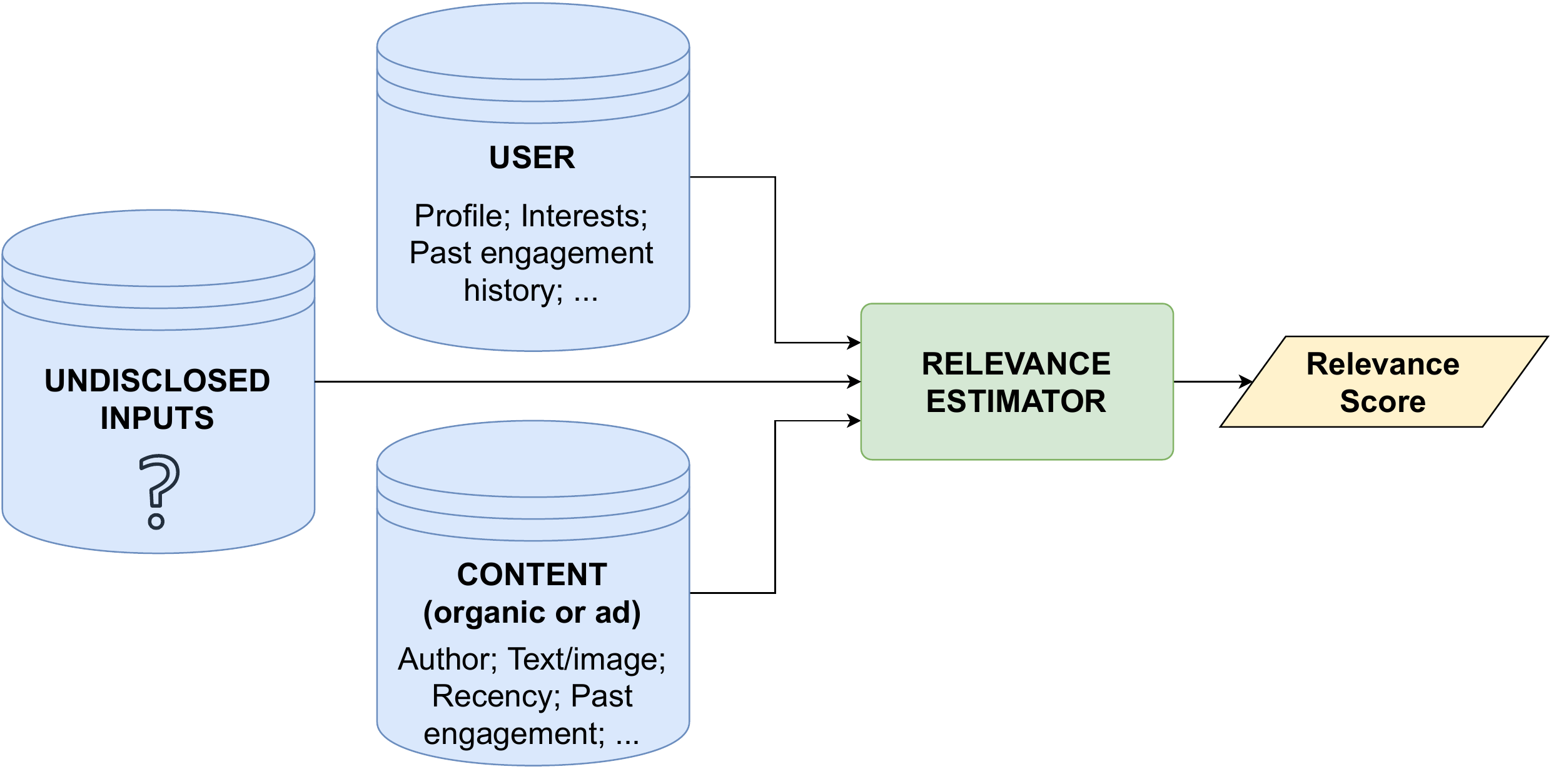}
    \caption{Social media platforms use relevance estimators to score
     every content using many factors as input, only some of which are publicly documented.}
  \label{fig:relevance_score}
\end{figure}

Given the vast amount of potential content shared on social media,
    relevance estimators have become responsible for selecting which
    content is shown on a user's feed and in what order, and which is omitted or deprioritized.
For example, for organic content, it may be selecting and ranking posts from
    a user's friends or pages they follow, or for promoted content,
    it may be running auctions for ads that are currently competing for a particular user's attention.
Platforms may mix both organic posts
    and ads in the content feed.
Facebook's algorithmic newsfeed dates back to 2007~\cite{FacebookNewsFeedAge},
  and Twitter and Instagram deployed such personalization in 2016~\cite{AlgorithmTimeline2016}.
Before deploying these algorithms, the platforms
  used ordering that was reverse chronological.

For organic content,
  these algorithms ultimately boil down to \emph{relevance scores}
  that will determine the selection and order of
  content shown at the top of users' feed.
For example,
  Facebook makes a number of predictions about how likely
  a user is to engage with posts,
  and will ``add these predictions up into a \emph{relevancy score}''
   to order the posts~\cite{FacebookTimeline}.
Instagram follows a similar approach~\cite{InstagramTimeline2021}.
Similarly, Twitter uses a number of ``consolidated signals to develop a
  \emph{relevancy score}''~\cite{TweeterTimeline2017}
  that it uses to determine which tweets to show on top of the feed.
LinkedIn also uses an algorithm that
 ``\emph{scores} tens of thousands of posts and ranks the most
  \emph{relevant} at the top of the feed''~\cite{LinkedInTimeline2019}.
TikTok mixes content from both followed accounts and others
  using algorithms that optimize for
  ``effective relevance'' as a ``secret sauce''~\cite{TiktokRelevance}.

Relevance estimators are also used in ad auctions, that 
  consider relevance as a factor predictive of user engagement.
These predictions are combined with other factors,
  such as the bid and budget the advertiser set for the ad,
  to determine the auction winner.
Different platforms use different terminologies to refer to these predictions.
For example, Facebook, LinkedIn and Twitter refer to them as
  ``Estimated action rates''~\cite{FacebookAdAuction},
  ``Relevancy scores''~\cite{LinkedInAdAuction},
  and ``Quality scores''~\cite{TwitterAdAuction},
  respectively.
But they all have very similar purposes in that they are applied as modifiers to bids
  to determine which ad wins an auction.
Therefore,
  an ad with the highest bid may not
  win an auction if it is given a low relevance score by the algorithmic prediction.

Platforms provide little transparency into their relevance score algorithms,
  neither for organic nor promoted content.
As summarized in \autoref{fig:relevance_score},
  publicly available documentation gives a high-level description
  that platforms use information about
  the content itself,
  the author of the content, 
  and user's profile data.
However, the specific types of algorithms and inputs to those algorithms
  are not disclosed.
Facebook, Instagram and LinkedIn
  use thousands of factors to estimate relevance of
  posts~\cite{FacebookTimeline, InstagramTimeline2021, LinkedInTimeline2019}.
Similarly, Twitter's documentation shows they use advanced
  machine learning algorithms to predict relevance,
  where the ``list of considered features and their varied
  interactions keeps growing''~\cite{TweeterTimeline2017}.
A recently leaked source code of a Russian search engine
  Yandex revealed 1,922 different factors used to rank content
  (albeit many are marked as legacy)
  which gives an additional evidence that platforms
  use many hidden factors to rank content~\cite{YandexLeak}.

The importance of relevance estimators to organic content and ad delivery
  lead us to place them at the center of
  our mechanism for platform-supported auditing in \autoref{sec:platform_supported}.

\subsection{Relevance Estimators Can Cause Various Forms of Algorithmic Harms}
  \label{sec:relevance_harms}

We next present examples of audits
  where controlled experiments demonstrate
  biased or harmful outcomes that result
  from social media platforms' algorithmic choices of relevance estimators.

Ad delivery is an area where there is substantial evidence
  for relevance optimization resulting in bias and discrimination.
Starting with Sweeney's empirical study in 2013~\cite{Sweeney2013},
  researchers hypothesized that platform-driven choices
  result in discriminatory ad delivery across demographic
  groups.
In 2019, this hypothesis was confirmed by Ali and Sapiezynski et al.~by showing Facebook's relevance algorithms skewed delivery
  of job and housing ads by gender and race,
  even when an advertiser targets a gender- and race- balanced
  audience and market effects are accounted for~\cite{Ali2019a}.
A subsequent study by Imana et al.~controlled for job
  qualifications, a legally excusable source of skew, to demonstrate
  that Facebook's relevance algorithms may be violating
  U.S. anti-discrimination laws~\cite{Imana2021}.
These examples provide evidence
  that opaque optimization algorithms
  result in discriminatory delivery of opportunity ads
  for certain demographic groups.
  
Facebook's academic work~\cite{do2022online} and Facebook's public statements~\cite{FacebookvsHUD} in response to the recent settlement with the US Department of Justice~\cite{FacebookvsHUD2} both also acknowledge the need to ensure its algorithms for opportunity ads are not biased. The legal developments serve as additional evidence that harms
  of relevance estimators that prior studies
  pointed out are well grounded.

Delivery of organic content is another area
  where past audits have found evidence for bias.
Twitter conducted an internal audit on its algorithms
  used to curate timelines, and found that its platform amplifies
  right-leaning political tweets more than moderate ones~\cite{TwitterAlgAmplification}.
The study suggests the difference in the amplification may be attributable to Twitter's ranking models
  assigning higher relevance scores to the right-leaning tweets.
Another external audit by Bartley et al.~performed a more general comparison
  of algorithmic and reverse chronological timelines on Twitter and
  showed the algorithmic timeline distorts information that users are shown~\cite{Bartley2021}.
  
 Besides such internal and external audits,
   investigations done by journalists have corroborated
   that the potential harms of algorithms used by platforms
   are not merely theoretical.
A recent prominent example is ``the Facebook files'', an investigation by
  Wall Street Journal into leaked internal Facebook documents.
Among other findings, the investigation showed how changes in 2018
  to make ``the platform healthier'' by focusing on
  relevance and engagement caused its algorithms to promote
  objectionable content~\cite{WsjFacbookFiles2021}.
They report cases where Facebook's algorithms
  have led teenagers to harmful content~\cite{WsjFBFilesTeens2021} and
  spread hateful posts~\cite{WsjFBFilesHateful2021}.
  
These findings underscore the need for public oversight.
The goal of such oversight will be to ensure relevance estimators
  that optimize for business objectives take societal interests into account.
Platform-supported auditing is an important part of progress towards
  this goal.

\section{Privacy-preserving Platform-Supported Auditing}
  \label{sec:platform_supported}

We next describe our approach to platform-supported auditing
  and how it addresses risks to the privacy of users and business
  interests of platforms.

\begin{figure}
  \centering
  \includegraphics[width=0.75\columnwidth]{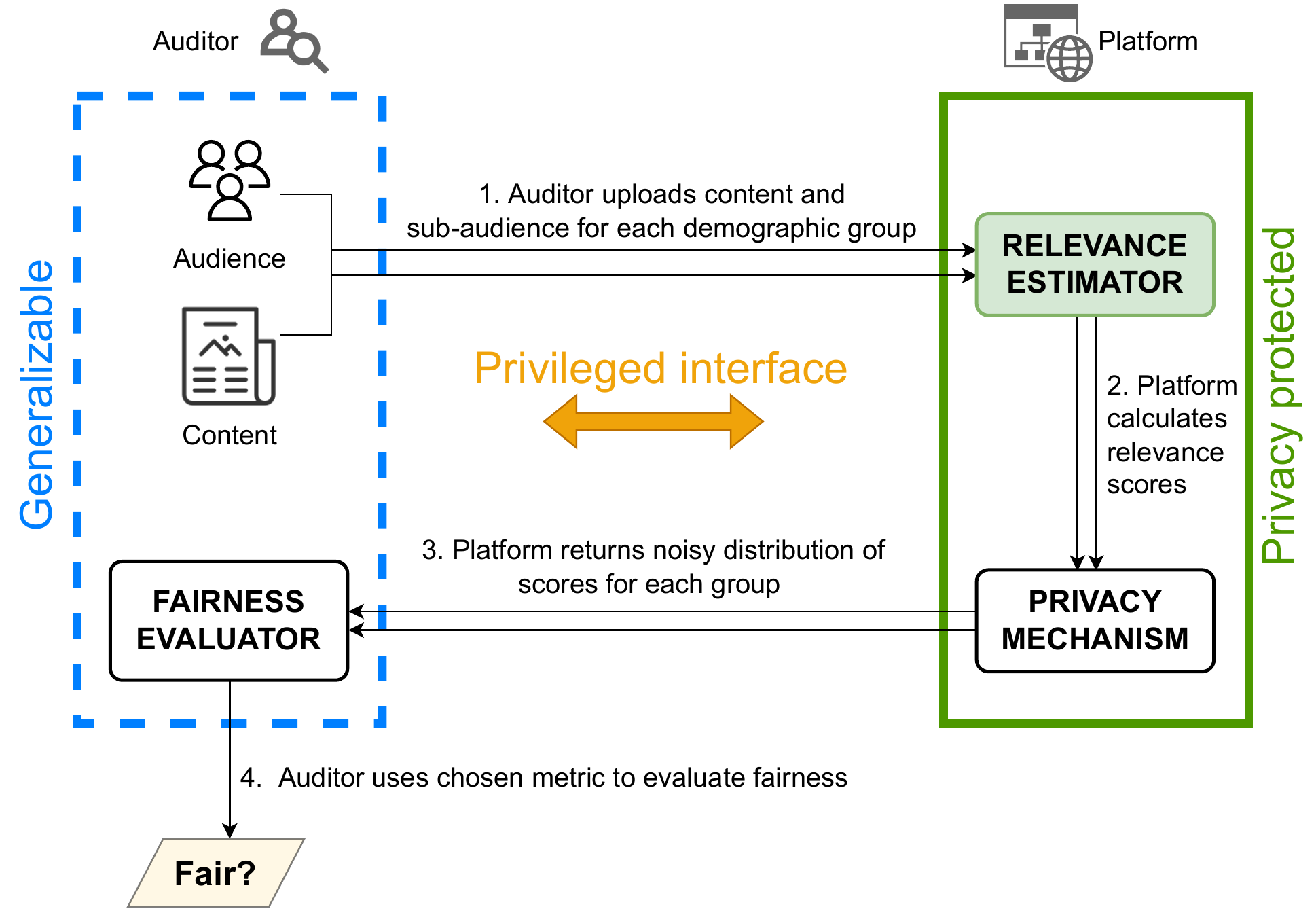}
    \caption{Privacy-preserving platform-supported auditing framework for evaluating fairness of relevance estimators.}
  \label{fig:auditing_overview}
\end{figure}

\subsection{Overview and Context}
  \label{sec:overview_framework}
  
Our proposal for \emph{platform-supported auditing}
  allows an auditor to evaluate whether,
  for a given piece of content,
  the platform's relevance estimator scores that content
  with bias reflecting protected attributes
  such as gender or race.
The framework is
  summarized in~\autoref{fig:auditing_overview} and has four high-level steps:
  (1) an auditor selects a trial content and an audience
  whose demographic attributes are known to the auditor,
  and uploads the content and sub-audience for each demographic group separately;
  (2) for each demographic group, the platform calculates relevance scores that estimate
  how relevant the content is to each user in the group;
  (3) the platform then applies a privacy mechanism
  and returns to the auditor a noisy distribution of the scores
  for each group;
  (4) finally, the auditor evaluates the fairness of the scores assigned
  to different demographic groups using an applicable metric of fairness.
We discuss each of these steps in more detail in~\autoref{sec:steps_details}.

\subsubsection{Key Properties of the Framework}  \reviewfix{CSCW A1, B1, F3: Describe first the high-level properties of framework before
 going into technical details to make the key takeaways digestible by CSCW audience.}
{ \MarkChange
Our framework has three key properties:
  privileged access,
  generalizability,
  and preservation of privacy.
\autoref{fig:auditing_overview} shows the relationship between
  these properties,
  and we next explore them
  to provide context for the details
  of our framework.  

\textbf{Privileged interface:}
A unique property of our framework is that we propose
  platforms provide to auditors
  a new, privileged query-access to the platform's relevance estimators.
This access is privileged in that it will be available only to auditors,
  not regular users or advertisers.
As summarized in~\autoref{sec:relevance_brain},
  relevance estimators estimate
  the interest a user is expected to show for each particular content.
 Given the central role of these algorithms to platform business models,
      platforms hold them closely and they are opaque to both users and advertisers. 
The framework we propose will increase transparency by
  giving auditors a privileged interface to
  query these algorithms,
  allowing evaluation of the algorithms' (un)fairness or potential harms
  while retaining privacy of both user data and platforms'
  proprietary information.
  
\textbf{Generalizable:} 
Our framework generalizes
  to allow
  the auditor to vary the audience, content, and fairness metric
  to evaluate many potential scenarios for bias.
This property is shown in the left hand side
  of~\autoref{fig:auditing_overview} (blue, dotted box).
For example, an auditor may evaluate performance of relevance estimators on job ads,
  political content, news, or misinformation to consider
  different societally relevant questions.
The auditor can select the audience to consider bias relative to
    attributes such as race, gender, age, or political affiliation.
The content and audience then
  determine what fairness metric the auditor uses as 
  what is considered ``fair'' is context-dependent~\cite{Verma2018}.
Our approach is generalizable because it allows auditors to explore
  multiple combinations along these three axes,
  and study various types of potential algorithmic harms.
  
\textbf{Preservation of privacy}:
The final important property of our framework is that it is privacy-preserving.
Our proposal keeps both user data and platform algorithms private from the auditor,
  as shown on the right side in~\autoref{fig:auditing_overview} (green, solid box).
We protect user information with Differential Privacy,
  and platform's algorithms and any proprietary information
  is protected by providing only query-level access.
We also minimize
  the chance of abuse by limiting query access to researchers
  vetted through the legal framework described in DSA and PATA.
}

\subsubsection{Comparison with Existing Taxonomy of Methods}

Our platform-supported auditing framework
  is a new category
  that extends the classical Sandvig taxonomy of auditing~\cite{Sandvig2014}.
Their taxonomy defines five types of audits:
\reviewfix{}
{ \MarkChange
  source code audit, noninvasive user audit, scraping audit, sock-puppet audit and crowdsourced audit.
}
Our platform-supported auditing defines a new type of audit because,
  unlike methods that use only features or APIs publicly available to regular users and advertisers,
   it requires a privileged and auditor-specific interface.
And unlike methods that require access to
  platforms' proprietary source code,
  it only requires query-level access to the output of their algorithms.
We further discuss how platform-supported auditing
  compares to the existing taxonomy of methods in~\autoref{sec:related_work}.

 \reviewfix{}
{ \MarkChange
Our framework is also different from collaborative user-driven audits,
  another class of methods that does not fit into the Sandvig taxonomy~\cite{Shen2021, DeVos2022}.
In particular,
  our proposal for direct query access to relevance estimators
  can help uncover biases that otherwise would be
  difficult to systematically study solely based on what content
  end-users' are shown shown on their newsfeed.
We also further discuss how our framework compares to user-driven audits in~\autoref{sec:related_work}.
}
  
\subsection{Privacy and Business Risks of Platform-Supported Auditing}
  \label{sec:why_privacy}

Our approach is designed to minimize risks to the privacy
  of platform users and to platform's proprietary information.
As discussed in~\autoref{sec:policy_push},
  protecting against these risks is an important goal
  of PATA and DSA,
  and is also a concern that platforms identify as
  a constraint to enabling transparency and auditability.
We next discuss the potential risks of
  providing query access to relevance estimators and
  the need for ensuring rigorous privacy protection when their outputs are shared.

 Relevance scores may leak private user data
   based on which they are calculated.
As discussed in \autoref{sec:relevance_brain},
  platforms calculate relevance scores based on
  users' personal profile data and
  their historical engagements with the platform.
The relevance of each particular content to each user
  may reveal information about the user that
  the auditor otherwise would not know.
For example,
  when a platform finds content about
  disability support or insurance highly relevant to a
  given user, that result suggests the user may
  be disabled or is caring for a disabled person.
Similar real-life examples from other contexts include
  Target's predictive algorithms for sending relevant coupons
  leaking a teenager's pregnancy~\cite{TargetPregnancy},
  and Facebook's ``People You May Know''
  feature that suggest connections Facebook deems relevant
  revealing private information about a user~\cite{FBPeopleYouMayKnow}.  
Even if relevance scores are aggregated in some fashion,
  prior work has shown similar aggregate outputs
  of personalization systems,
  combined with auxiliary information about users
  can leak private information~\cite{Calandrino2011, Weinsberg2012, Beigi2020}.
Therefore, our auditing method must limit the
  potential to make such inferences.

In~\autoref{sec:add_dp_noise},
  we show how our framework protects the privacy of users,
  when privacy protection is defined as ensuring a Differential Privacy (DP)
  guarantee on any \emph{data that is shared with the auditor}.
\reviewfix{}
{ \MarkChange
Platforms often cultivate information about users,
  including sensitive information,
  but our goal is to ensure that no such information becomes available to the auditors.
(addressing risks to privacy that the platform itself may pose is outside the scope of our paper.)
}
DP is the current gold standard
  for protecting privacy of individuals,
  while providing useful statistical computations on their data~\cite{Dwork2006}.
DP provides a rigorous guarantee
  that each individual's choice to participate in the audit
  has negligible impact on their privacy.
A differentially-private mechanism will specifically protect 
 the privacy of \emph{the users participating in the audit},
 while providing aggregate information
 about the relevance estimator, which
 the auditor can use to asses fairness.

In addition to risks to platform users,
  platforms themselves would like
  to minimize what details of their algorithms they share.
\reviewfix{}
{ \MarkChange
Platforms regard their algorithms' source code and data
  as proprietary business assets;
  our framework explicitly does not require direct access to either.
 } 
Our framework minimizes information it requires the platforms to share about their algorithms and data
  by providing only query access to auditors,
  asking to share only aggregate relevance metrics,
  while preserving the confidentiality of the source code, how those metrics are computed and
  what inputs and training data they use.

\subsection{Steps of Platform-Supported Auditing}
  \label{sec:steps_details}
  
We next describe each of the four steps of platform-supported
  auditing in detail.
\subsubsection{Auditor Uploads Content and Audience}
  \label{sec:auditor_uploads}

The auditor first will select a trial content
  and a customized audience.
The auditor selects the content and demographics based on
  the specific platform and the type of algorithmic harm
  they are studying.
The audience is a list of users whose demographic attributes are known
  to the auditor.

The auditor then uploads the content and a
  sub-audience for each demographic group to the platform.
Major platforms already have an infrastructure
  for advertisers to upload audience and content which,
  with some modifications,
  can be used for auditing purposes.
\reviewfix{}
{ \MarkChange
Given our goals of protecting individual users
  from harm, 
  we focus on accounts of typical users.
Commercial accounts, or accounts of unusual users such as
  public figures or prominent influencers, are outside the scope of our work.
}

The type of content and the demographic make up of the users depends
  on what type of harm or bias the auditor is interested in studying.
For example, one may wish to study whether LinkedIn's ad delivery algorithms
 deliver STEM job ads in a biased way that reflects
  historical under-representation of women in the field.
To perform the study,
  the auditor may use a STEM job ad as the content,
  and a sample of men and women as the audience.
The auditor will query the platform using the audience
  from each demographic group
  to evaluate whether LinkedIn's relevance estimators
  assign higher scores to men compared to women for the STEM job ad.
 
\reviewfix{}
The auditor will specify the audience by uploading a custom
  list of specific people whose demographic attributes are
  known to the auditor.
{ \MarkChange
The auditor can build such an audience in two ways: by externally recruiting
  volunteers for the study or by asking the platform to provide a random sample of users. }
Recruiting volunteers has two main advantages over prior methods
  that use publicly available datasets, such as voter data~\cite{Ali2019a, Ali2019b, Imana2021}.
First, each participant gets the opportunity to consent to the use
  of their data for auditing purposes.
 Second, participants can provide additional attributes that
  may not be present in public data but could be useful for auditing.
An example is a job qualification attribute that is useful for auditing
  delivery of employment ads.
Existing recent studies such as The Markup's Citizen Browser~\cite{CitizenBrowser}
  and Mozilla's Rally~\cite{MozillaRally} show that users
  are willing to opt-in and provide data to reputable efforts that aim to hold platforms accountable.

{ \MarkChange
In addition to such external efforts, platforms themselves can provide support for
  selecting a sample of users among those who have provided the platforms with their
  sensitive attributes~\cite{metatech, LinkedInSelfID}.
Using platforms' support for building a random audience of users can help minimize
  bias such as self-selection that can be present when
  recruiting external volunteers.
We discuss the potential limitations due to audience selection approaches
  in~\autoref{sec:assumptions}.}
  
Having the auditor specify a custom audience is also
  advantageous over letting the platform itself pick an audience.
It helps protect the privacy of users since it does not require platforms,
  nor gives them the excuse,
  to collect sensitive demographic attributes.
This advantage also addresses the challenges around
  collecting and securing sensitive attributes of users
  that companies often identify as
  one of the major obstacles
  to auditing for fairness~\cite{Bogen2020, Andrus2021, metatech}.

Popular social media platforms
  have an infrastructure for advertisers to upload
  custom audiences.
These existing features are currently designed for use by
  advertisers to run ad campaigns
  that retarget their customers.
An advertiser may upload information such as
  names, email addresses and phone numbers
  of their customers and the platform tries to match
  this information with user profiles on the platform.

These existing features can serve as a starting point
  for platforms to build a similar interface that can be
  used for platform-supported auditing.
In the existing custom audience features, 
  not all people in the audience
  may match to user profiles on the platform.
In a prior work,
  we have found that such partial matches can be a
  source of error when using custom audiences for auditing~\cite{Imana2021}.
A possible modification for supporting accurate auditing is for the
  platform to allow the auditor to upload unique identifiers
  (for example: Facebook usernames)
  for the accounts of people who are participating in the audit.
The auditor can collect these identifiers when recruiting
  volunteers for the study.

\subsubsection{Platform Calculates Relevance Scores}

The platform then calculates how relevant the content is to each user
  in the custom audience.
Relevance estimation on the platform boils down to a~\emph{relevance score}
  for each user,
  which is the platform's prediction of how likely the user is to engage
  with the content.
The platform will not report the raw scores to the auditor as they may reveal
  private information about its users' past engagement history.
Instead, the platform builds statistics that
  summarizes the distribution of the scores (for example, a histogram or a CDF),
  and adds privacy protections (discussed in \autoref{sec:add_dp_noise}),
  before returning the statistics to the auditor.

Platforms use many factors for estimating relevance
  but not all are applicable for auditing context.
For example, during a normal usage of the platform,
  the estimators may use as inputs factors such as
  what time of the day it is and
  for how long a user has been logged in~\cite{FacebookTimeline}.
For such temporal variables that are only
  applicable in the context of a user
  browsing the site,
  the platform must keep them constant for all users in the audience.
This control allows the auditor to evaluate
  bias in the relevance estimators that may arise from the historical
  data that the platform has about users and not
  from temporal factors.

\subsubsection{Platform Applies Privacy Mechanism and Returns DP-protected Scores}
  \label{sec:add_dp_noise}

The platform then applies a differentially private mechanism
  to the statistic of relevance scores calculated
  and returns a noisy statistic to the auditor.
The mechanism will provide a rigorous guarantee that
  the data the auditor gets 
  was produced in a way that ensures differential privacy for
  individuals participating in the audit.
We use the following definition of DP, where
  neighboring databases are defined as differing in one person $u$'s data:
  $D_1 = D$ and $D_2 = D \cup \{u\}$ for some database $D$.
\begin{definition}[$\epsilon$-Differential Privacy~\cite{Dwork2006}]\label{def:dp}
Given a privacy parameter $\epsilon > 0$, a randomized mechanism $M$ is $\epsilon$-differentially private if
  for any two neighboring databases $D_1$ and $D_2$,
  and for any subset $S \in R$ of outputs,
$$Pr[M(D_1) \in S] \le e^{\epsilon} * Pr[M(D_2) \in S],$$
where the probability is taken over the random coin tosses of $M$.
\end{definition}

Auditors can approximate tests for group-fairness metrics using a
  binned histogram of relevance scores without access to
  individual scores.
One method to share the binned histogram while preserving privacy is
  using the Laplace Mechanism~\cite{Dwork2006}.
The platform can independently add noise drawn from the
  Laplace distribution to each of the bins
  in the histogram.
Since presence or absence of a single user changes
  each bin's count by at most one,
  adding noise from the Laplace distribution with scale $1/\epsilon$
  independently to each bin ensures the mechanism
  is $\epsilon-$differentially private~\cite{Dwork2006}.
The platform then returns the noisy histogram counts back
  to the auditor.
 \reviewfix{}
{ \MarkChange
At the end of~\autoref{sec:main_result}, we discuss
  what choices of the privacy parameter $\epsilon$ our framework allows.}

\reviewfix{}
{ \MarkChange
We describe above one iteration of an audit but, in reality,
  an auditor may be interested in using the same audience
  to study multiple questions.
Answering different questions may require querying relevance
  estimators multiple times, where each query
  uses up additional privacy budget. 
For such cases,
  (a well-studied topic in the DP literature) composition property of DP allows for the auditor
  to split a total privacy budget among the different
  queries~\cite{dwork2014algorithmic}.
We leave exploration of how such total budget can be best  allocated
  and at what cadence it can be replenished as an area
  of future work.
}

\subsubsection{Auditor Evaluates Fairness of Relevance Scores}
  \label{sec:eval_fairness}

Finally, the auditor uses the noisy distribution of scores
  to test whether there is a disparity between the relevance
  scores the algorithm assigns to the different demographic groups.
Any arbitrary post-processing to
   the output of the differentially private mechanism from the prior step
   does not reverse the privacy protection.
Therefore, the auditor can use the noisy scores to apply
   any post-processing computations to test for fairness without
   further reducing the privacy of the users.  
  
The specific metric of fairness depends on the type of algorithmic
  bias the auditor is interested in testing for.
For example,
  to study bias in the delivery of employment ads,
  the auditor may use Equality of Opportunity as a metric
  for fairness, since it takes qualification
  of people for jobs into account, which is a relevant factor for the context
  of employment~\cite{Hardt2016}.
We further explore this scenario in our theoretical result in~\autoref{sec:privacy_tradeoffs}.

\subsection{Trust Model and Limitations}
	\label{sec:model_limitations}

We next discuss the trust model we use
  to evaluate the privacy and business interest risks of our approach.
The efficiency our approach assumes a legal framework in
  which both the platforms and auditors work in good faith.

\textbf{Platforms:} One major assumption of our framework is
  that the platform will truthfully collaborate
  with auditors and ensure audits are done
  accurately and effectively.
The platform must provide auditors access to
  the same algorithms that are used in production,
  truthfully executing them on the audience the auditors upload
  and reporting relevance scores accurately (modulo privacy modifications).
This assumption was not stated in prior auditing methods
  that do not use a platform's support.
Even for such methods,
  platforms have the means to know they are being
  audited as the audiences and methodologies auditors typically used
  are publicly documented.
Examples included North Carolina's voter datasets used
  as data source for demographic attributes~\cite{Speicher2018},
  Facebook ad accounts used to audit ad delivery~\cite{Ali2019a, Ali2019b, Imana2021},
  and browser extensions used
  for collecting data from Facebook~\cite{NYUAdobservatory, CitizenBrowser}.

Assuming the platform truthfully collaborates with auditors is a strong assumption,
  but there are four reasons we think it is appropriate.
First, 
  the consequences of non-compliance are significant when
  auditing is part of an official legal framework,
  as it would be in the context of a DSA- or PATA-like law
  or a legal settlement, such as Facebook's settlement with the US Department of Justice~\cite{FacebookvsHUD}.
For example,
  Volkswagen faced significant legal and financial
  repercussions as a result of their violation of emissions regulations~\cite{Jung2019}.

Second, platforms also have the incentive to minimize
  inadvertent errors in order to avoid tarnishing their public image
  and potential legal liability.
Two cases, both involving Facebook,
  serve as an example of this incentive.
In the first case,
  Facebook made inadvertent errors in sharing data
  to external researchers as part of its Social Science
  One program~\cite{FBSocialOneData}.
This preventable error undermined academic work
  that was based on the data~\cite{FBSocialOneData},
  tarnishing Facebook's efforts to be a leader in increasing transparency.
In the second case,
  Facebook mistakenly inflated potential reach estimates
  for ads,
  and was sued as a result~\cite{FBInflatedAdMetrics}.

Third, simply formalizing auditing and involving two
  parties often adds sufficient oversight to discourage
  abuse.
For example,   
  corporate financial accounting is not immune to fraud,
  but the levels of non-compliance are small enough that
  it is a very useful and powerful tool.
  
Finally,
  as discussed in~\autoref{sec:beginning-to-favor},
  there is evidence that the platforms themselves may be moving
  towards supporting audits through giving external
  researchers privileged access to their data and algorithms.

\textbf{Auditors:}
The platform must also trust researchers
  doing the independent audit.
One risk for abuse is misuse of the auditing interface to harm
  a platform's business.
Both the DSA and PATA provide rules to ensure
  only vetted researchers
  will be allowed to perform audits on social
  media platforms~\cite{Pata2022, dsa2021}.
In both proposals, an assigned regulatory body will screen
  researchers and their projects before they are allowed to audit
  a platform’s system or data.
Platform-initiated transparency efforts such as Facebook's FORT, Social Science One, and YouTube's Researcher Program also all have approaches for vetting researchers~\cite{SocialScienceOne, FacebookFORT, YoutubeResearchProg}.
Such screening processes will minimize the risk
  that comes from malicious auditors, and the platforms' implementations show that the platforms themselves believe this risk can be overcome.

Another risk is misuse of sensitive data that auditors collect
  from users who are participating in an audit. %
\reviewfix{}
{ \MarkChange
Because DP protects user information that may leak to auditors,
  our work does not create new privacy risks to users.
As one example, the auditor framework does not create new opportunities that
  allow governments or other third parties to surveil users.
}
In addition,
 unlike prior methods that used voters' data without their knowledge,
  users in our proposed framework would voluntarily participate
  by being recruited by an auditor (as discussed in~\autoref{sec:auditor_uploads})
  or through programs the platforms provide~\cite{LinkedInSelfID}.

\section{Sample Size Required for Auditing Relevance Estimators with Privacy}
  \label{sec:privacy_tradeoffs}
  
We next present the key technical result of this paper by applying our framework to one
  use-case: a study of discrimination in employment ad delivery.
We show that the addition of differential privacy
  to the auditing pipeline does not prevent an auditor from achieving
  the same statistical confidence as without privacy protections,
  provided the sample audience is increased by a small constant factor.
This result supports our claim that
  it is feasible to both audit for fairness
  \emph{and} protect user privacy and platforms' business interests.

\subsection{Setup and Assumptions: Bias in Delivery of Employment Ads}
  \label{sec:setup}
Auditing social media platforms for fairness while preserving privacy
 is a goal that desirable in multiple scenarios.
We study one scenario: assessing discrimination in delivery of employment ads.
Our problem formulation is general,
  although specific scenarios
  place additional requirements,
  like the role of job qualifications in employment ads.
Extending our approach to other types of ads
  may require identifying similar factors reflecting allowable preferences.

We consider the case where an auditor wishes to confirm
  delivery of job ads is unbiased relative to a factor
  such as gender or race.
To evaluate this question,
  the auditor will examine the relevance scores
  a platform's relevance estimator will assign
  to different groups with specific demographic attributes.
This scenario is motivated by prior third-party audits that have
  indirectly measured the role of relevance optimization in
  biased job ad delivery~\cite{Ali2019a, Imana2021}.

\subsubsection{Setup and Definitions:}
We first introduce formal notations for the scenario.
Let  $X$ represent a set of all users on a platform and
  let $A$ be the range of values for a sensitive attribute
  (For example, $A$ =  $\{$black, white, ...$\}$ for race).
Let $Q=\{0,1\}$ represent binary options for qualification of a user
  to a given job ad (1 if the user is qualified, 0 -- otherwise).
Let $R_j(x)$ be the relevance estimator that calculates the
  relevance score of the job ad $j$ to a given user $x \in X$.
We assume a specific ad $j$ and omit the subscript $j$ throughout.
And let $Y$ be a small finite set of discrete relevance scores
(we describe how to extend $Y$ to the continuous case at the end
  of this section).

In practice, the external auditor cannot have access to a complete list of all of the platform's
  users ($X$), so the auditor recruits a sample ($S$) of users to perform the audit.
The auditor uses a random sample set
  $S = \{(x_1, a_1, q_1), (x_2, a_2, q_2), ...., (x_n, a_n, q_n) \}$
  drawn i.i.d.~from $X$.
In that case, each subset $S_{a,q}$ is also i.i.d.~in $X_{a,q}$,
  where $S_{a,q}$ and $X_{a,q}$ represent subsets with given values of $a$ and $q$.
We discuss implications of this assumption at the end of this section.

Following the steps in \autoref{fig:auditing_overview},
 the auditor first queries the platform's relevance estimator
  using each subset $S_a$ and ad $j$ (step 1).
The platform then applies $R$ to every user in $S_a$ (step 2)
  and builds a histogram $H$ of the scores,
  grouped by possible range of relevance scores in $Y$.
It then independently adds noise drawn with a
  Laplacian distribution $Lap(\frac{1}{\epsilon})$ to each of the bins in $H$,
  where $\epsilon$ represents the level of differential privacy desired.
The platform returns the noisy histogram counts back
  to the auditor (step 3).

Finally, the auditor tests for fairness of the scores assigned
  using Equality of Opportunity as a definition of fairness (step 4).
Equality of Opportunity is an established fairness notion in the algorithmic fairness literature, and is applicable to job ads
  as it allows for taking into account the qualification of users~\cite{Hardt2016}.

 \begin{definition}[derived from Equality of Opportunity~\cite{Hardt2016}]\label{def:eoo_def}
A relevance estimator function $R$ satisfies equality of opportunity:
  \begin{align*}
  & Pr_{(x, a', q) \sim\ X}[R(x) = y | a' = a \land q = 1] \\
  &= Pr_{(x, a', q) \sim\ X}[R(x) = y | q = 1]~\textrm{for all}~a \in A~\textrm{and}~y \in Y,
  \end{align*}
where the probability is taken over the choices of samples from $X$ and the random coin tosses of $R$.
\end{definition}

We modify Hardt et al.'s formulation by using
  the group of qualified people ($q=1$) to represent the ``advantaged outcome'' group~\cite{Hardt2016}.
The advantaged outcome in our case is that a person sees a job ad
  because they are qualified for the job.
In addition, in our formulation, the outcome space $Y$ is not binary but
  a finite set of discrete values.

To test for this metric, the auditor must know whether
  each user is qualified for the job being advertised.
For convenience, we introduce the following notation:
\begin{equation}\label{eq:true_P_a_r}
P_{a, y}(R) = Pr_{(x, a', q) \sim\ X} \left[  R(x) = y | a' = a  \land q = 1  \right]
\end{equation}
$P_{a, y}(R)$ represents the likelihood that a qualified individual from
  a specific demographic group $a$ receives a relevance score $y$.
The auditor expects this likelihood to be equal across demographic groups if the platform's algorithm is unbiased.
 
We relax strict equality of the above term since any real-world
  observation may have small noise or variation.
We will use a relaxation from prior work~\cite{Segal2021},
that allows a small additive error $\alpha$ 
  as maximum allowed fairness gap ($FG$) between any two demographic groups.
We change the relaxation to use $\alpha$ instead of $\epsilon$ because we use
$\epsilon$ as a privacy parameter.

\begin{definition}[$\alpha$-fairness~\cite{Segal2021}]\label{def:alpha_fair}
We define a relevance estimator function $R$ to be $\alpha$-fair if:
$$FG(R) = \max_{a_1, a_2 \in A, y \in Y} |P_{a_1, y}(R) - P_{a_2, y}(R)| \le \alpha $$
\end{definition}

Since the auditor has only access to
  an independent sample of users ($S$),
  the measure of $P_{a,y}$ the auditor gets empirically is given by:

\begin{equation}\label{eq:p_a_y}
\overline{P}_{a, y}(R,S) = \frac{1}{n_{a,q}} \sum_{i=1}^{|S|} \mathbb{1}\{ R(x_i) = y \land a_i = a  \land q_i = 1 \}
\end{equation}

where  $\mathbb{1}\{ .\}$ is an indicator function selecting qualified members from group $a$ that are assigned a score $y$,
and $n_{a,q}$ is the number of qualified members in S
from group $a$.
The equation requires that $n_{a,q}>0$, an assumption we discuss at the end of this section.
  
Let $n_{a,q,y}$ be the number of
  qualified people in $S$ from group $a$ that got assigned
  a score $y$. We can also rewrite $\overline{P}_{a, y}(R,S)$ as:

\begin{equation}\label{eq:n_a_q_y}
\overline{P}_{a, y}(R,S) = \frac{ n_{a, q, y}}{n_{a, q}}
\end{equation}

We next consider the value of $\overline{P}_{a, y}(R,S)$ after it is distorted by noise to preserve privacy.
From \autoref{eq:n_a_q_y},
  $n_{a, q}$ is already known to the auditor so the quantity the platform wishes to protect is $n_{a,q,y}$,
  which represents each bin in the histogram that the platform computes.
The platform applies Laplace mechanism by adding noise drawn from
$r ~ \sim Lap(\frac{1}{\epsilon})$
to each count $n_{a,y,q}$ to guarantee $\epsilon$-DP~\cite{Dwork2006}.
Let $P^{*}_{a, y}(R,S)$ represent the noisy value the platform calculates:

\begin{equation}\label{eq:p_a_y_noisy}
P^{*}_{a, y}(R,S,\epsilon) = \frac{ n_{a, q, y} + r}{n_{a, q}} = \overline{P}_{a, y}(R,S) + \frac{r}{n_{a, q}}
\quad \textrm{s.t.} \quad r ~ \sim Lap(\frac{1}{\epsilon})
 \end{equation}

Extending a formulation in prior work~\cite{Segal2021}
  by adding a new privacy parameter,
  the empirical fairness gap (EFG) is given below
  (we give both the private and non-private cases).
A large EFG between two demographic groups implies unfairness.
$$EFG(R, S) = \max_{a_1, a_2 \in A, y \in Y} |\overline{P}_{a_1, y}(R,S)- \overline{P}_{a_2, y}(R,S)| $$
$$EFG(R, S,\epsilon) = \max_{a_1, a_2 \in A, y \in Y} |P^{*}_{a_1, y}(R,S,\epsilon)- P^{*}_{a_2, y}(R,S,\epsilon)| $$

The auditor checks $EFG(R,S,\epsilon) \le \alpha$ to test whether a relevance estimator $R$ is fair.
To analyze the sample size needed to perform this test with high statistical confidence,
  we will use the following definition that allows a small $\delta$ probability of failure over the
  randomness in $R$ and possible choices of samples in $S$.
  
\begin{definition}[($\alpha, \delta)$-fairness~\cite{Segal2021}]\label{def:alpha_delta_fair}
We define $R$ to be $(\alpha, \delta)$-fair with high probability with respect to S if:
\begin{align*}
& Pr \left[EFG(R,S) \le \alpha \right] =  Pr \left[ \max_{a_1, a_2 \in A, y \in Y} |\overline{P}_{a_1, y}(R,S) - \overline{P}_{a_2, y}(R,S)| \le \alpha \right] > 1 - \delta
\end{align*}

\end{definition}

We extend this definition for the case where an $\epsilon$-DP mechanism is applied to outputs of $R$
  to protect privacy of users.
  
\begin{definition}[($\alpha, \delta, \epsilon)$-fairness]\label{def:alpha_delta_eps_fair}
We define $R$ to be $(\alpha, \delta, \epsilon)$-fair with respect to S
  where an $\epsilon$-DP mechanism is applied to outputs of $R$ if:
  \begin{align*}
  &Pr \left[EFG(R,S, \epsilon) \le \alpha \right] =  Pr \left[ \max_{a_1, a_2 \in A, y \in Y} |P^{*}_{a_1, y}(R,S,\epsilon)- P^{*}_{a_2, y}(R,S,\epsilon)| \le \alpha \right] > 1 - \delta
  \end{align*}
\end{definition}
 
The formulation in this and the following sections assumes $Y$ is a set of discrete values.
\autoref{eq:true_P_a_r} and~\autoref{eq:p_a_y} can be extended to the case where $Y$
  is a continuous space
  by choosing a different indicator function and comparing CDFs of relevance
  scores:
  \begin{equation}\label{eq:p_a_y_continuous}
  \begin{aligned}
 &P_{a, y}(R) = Pr_{(x, a', q) \sim\ X}[R(x) > y | a' = a \land q = 1] \\
&\overline{P}_{a, y}(R,S) =  \frac{1}{n_{a,q}} \sum_{i=1}^{|S|} \mathbb{1}\{ R(x_i) > y \land a_i = a  \land q_i = 1 \}
 \end{aligned}
\end{equation}

\subsubsection{Assumptions:}
  \label{sec:assumptions}
\reviewfix{}
Our approach makes several assumptions to avoid degenerate cases.
We describe these next so that an auditor
  can design a robust experiment and
  may verify, post-audit, that the assumptions are met.

Equality of Opportunity (EoO) metric (Definition \autoref{def:eoo_def})
  adapts to unequal numbers of qualified individuals
  from different groups, but it cannot handle cases
  when \emph{no} one or \emph{very few} in the population
  with specific attributes are qualified for the job being advertised.
The first degenerate case occurs when $n_{a,q}=0$ in the denominator in~\autoref{eq:p_a_y}.
Another case is when only a few individuals are qualified from one group,
  and very many individuals are qualified from a second group (Example: $n_{a_1,q}=1$ and $n_{a_2,q}=1$ million).
In this case, EoO requires selecting all or none of the 1 million people in $a_2$ to match
  the inclusion or exclusion of the only individual in $a_1$.
Our \autoref{thm:main_result} guarantees that, for realistic parameters,
  $n_{a,q}$ is not small and that such degenerate cases do not occur.
Moreover, the auditor may verify, post-audit, that the assumptions about $n_{a,q}$ were met.

Second, we assume samples in each demographic group
  are independent and identically distributed.
We recognize that there maybe confounding factors
  that may induce bias,
  such as the location the audience is chosen from
  or difference in how active users are on the platform.
The auditor can anticipate some of these factors and
  control for them
  but only the platform has the data to verify independence.
In our result, we assume independence only within samples in a group,
  so we do not expect this limitation to decrease
  the observable differences in fairness across groups.
This assumption is common in nearly all statistical studies,
  and is aimed to be achieved by following best practices in subject selection.
\reviewfix{}
{ \MarkChange
Examples from prior work include repeating audits on various audience partitions,
  and varying locations that users are chosen from~\cite{Ali2019a, Imana2021}.
}

Third, we assume there is some way to randomly sample users.
This mechanism may be provided by the platform,
  or the auditor may use some external source of users
  (in which case we require that will not induce its own bias).
We recognize that sampling users from social media and
  encouraging them to share their data with the auditor may be difficult,
  but prior studies have met this requirement satisfactorily
  (for example, the work of Citizen Browser~\cite{CitizenBrowser}).
We therefore
  place this problem outside the scope of this paper.

\subsection{Result: Minimum Sample Size Required for Auditing with Privacy}
  \label{sec:main_result}

Building on the background in the prior section,
  we give the following theoretical result:
  we show that, for employment ad delivery use-case,
  auditing with differential privacy guarantee
  increases the number of samples required for auditing,
  but only by a small constant factor.

\begin{theorem}\label{thm:main_result}
An audit relying on a differentially privatized output of a relevance estimator $R$
  is ($\alpha, \delta, \epsilon$)-fair under equality-of-opportunity
  provided that, compared to the non-private case,
  an additional factor of $S_{dp}$ samples are measured.
We show that $4 \ln(3)/\ln(2) = 6.34$ is an upper bound for $S_{dp}$ and
  that ~4 is a better estimate for $S_{dp}$ under
  typical auditing parameters.

Formally, for an auditor to verify $R$ is $(\alpha, \delta, \epsilon)$-fair with respect to a sample set $S$,
  assuming $\epsilon > \alpha/2$,
  the condition $EFG(R,S, \epsilon) \le \alpha$ and the following
  condition on the minimum number of samples must hold:
\begin{equation}
\min_{a \in A} n_{a,q} \ge  \frac{8}{\alpha^2} \ln{\frac{3 |A| |Y|}{\delta}}
\end{equation}
where $S = \{(x_1, a_1, q_1), (x_2, a_2, q_2), ...., (x_n, a_n, q_n) \} \sim X$ 
and $n_{a,q}$ is the number of people in $S$ with sensitive attribute $a \in A$
and who are qualified for the job being advertised.
$\alpha$ and $\delta$ are knobs that control the level of fairness and statistical
confidence, respectively.
\end{theorem}

To prove this theorem,
  we first show with the case of auditing relevance scores
  when a privacy mechanism is not used.
We then analyze by what factor the required number of samples
  increases when a differentially private mechanism is applied.

\begin{lemma}\label{lemma:n_min_a_no_priv}
Without any guarantees of privacy,
  the following minimum number of samples is required
  to verify whether $R$ is $(\alpha, \delta)$-fair with respect to a sample set $S$:
\begin{equation}
\min_{a \in A} n_{a,q} \ge  \frac{2}{\alpha^2} \ln{\frac{2 |A| |Y|}{\delta}}
\end{equation}
where $S = \{(x_1, a_1, q_1), (x_2, a_2, q_2), ...., (x_n, a_n, q_n) \} \sim X$ 
and $n_{a,q}$ is the number of people in $S$ with sensitive attribute $a \in A$
and who are qualified for the job being advertised. 
\end{lemma}

For the non-private case, the proof directly follows from
   prior work by Segal et al.~on auditing
   machine learning models using cryptographic techniques~\cite{Segal2021}.
In \autoref{sec:non_private_proof},
  we extend their proof with consideration of qualification as an additional attribute.

We next consider sample size for the private case,
  where the auditor
  receives a noisy histogram of relevance scores because the platform applies
  a differentially-private mechanism.

\begin{lemma}\label{lemma:n_min_a_priv}
With privacy, 
  the following minimum number of samples is needed
  to verify whether $R$ is $(\alpha, \delta, \epsilon)$-fair with respect to a sample set $S$,
  \begin{equation}
\min_{a \in A} n_{a,q} \ge  \frac{8}{\alpha^2} \ln{\frac{3 |A| |Y|}{\delta}}
\end{equation}
where $S = \{(x_1, a_1, q_1), (x_2, a_2, q_2), ...., (x_n, a_n, q_n) \} \sim X$ 
and $n_{a,q}$ is the number of people in $S$ with the sensitive attribute $a \in A$
and are qualified for the job being advertised.
\end{lemma}

\begin{proof}
At a high level, the proof works by first defining a bad event that we
  want to happen with very low probability and
  then conditioning on this event not happening to derive the sample size needed to
  guarantee $(\alpha, \delta, \epsilon)$-fairness.
The  bad event is when there is error in the value for $P_{a,y}$ that the auditor calculates
  empirically.
We have two sources of error: sampling error
  and error due to noise added to protect privacy.

Now, consider the following ``bad'' event where the error between the value the auditor
  calculates $P^{*}_{a, y}(R,S)$ and the true $P_{a, y}(R)$ is above some threshold $t > 0$:
$$\text{``Bad''}:~ \left|P^{*}_{a, y}(R,S) -  P_{a, y}(R)\right| = \left|\left(\overline{P}_{a, y}(R,S) + \frac{r}{n_{a,q}}\right)-  P_{a, y}(R)\right| > t$$

Conditioning on the event that the total error for the bad event does not exceed $t$,
  we get a lower bound
  for a sample size that satisfies
  $(\alpha, \delta)$-fairness using the following value of $t$ (see~\autoref{sec:non_private_proof}):
\begin{equation}\label{eq:t_and_alpha}
t = \frac{\alpha}{2}
\end{equation}

We bound the probability of the
  above bad event for all groups in $A$ and possible outputs in $Y$: 
 $$Pr \left[\exists a\in A ~\text{and}~ y \in Y: \left|\left(\overline{P}_{a, y}(R,S) + \frac{r}{n_{a,q}}\right) -  P_{a, y}(R)\right| > t \right] \le \delta $$
 
By applying the triangle inequality,
  it is sufficient (but not necessary) to bound the probability that
 each of the two sources of errors exceed $t/2$:
\begin{equation}\label{eq:two_error_terms}
Pr \left[ \left|\overline{P}_{a, y}(R,S) -  P_{a, y}(R)\right| > \frac{t}{2} \right]  +  Pr \left[\left|\frac{r}{n_{a,q}}\right|  > \frac{t}{2} \right]
\end{equation}

Since we require the samples in $S$ are chosen i.i.d.,
  $\overline{P}_{a, y}(R,S)$ is unbiased estimator of ${P}_{a, y}(R)$,
  i.e, $E[\overline{P}_{a, y}(R,S)] = {P}_{a, y}(R)$ (We prove this in~\autoref{sec:hoef_with_bias}).
Therefore, we can apply Hoeffding's inequality to the first term (sampling error) to 
 simplify it to $2 \exp(\frac{- n_{a,q} t^2}{2})$.
  
We then apply a known tail bound
  for the Laplace distribution (for $r \sim Lap(B):~ Pr[|r| \ge t] < exp(\frac{-t}{B})$)
  to the second term (privacy error) to simplify it to 
$\exp(\frac{-n_{a,q} t\epsilon}{2})$.
We then take a union bound over all possible values of $a$ and $y$:

\begin{align*}
&Pr [\exists a\in A ~\text{and}~ y \in Y: \text{Bad event occurs}] \\
&\le \sum_{a \in A}\sum_{y \in Y} Pr \left[ \left|\left(\overline{P}_{a, y}(R,S) + \frac{r}{n_{a,q}}\right) -  P_{a, y}(R)\right| > t \right]  \\
&\le \sum_{a \in A}\sum_{y \in Y} Pr \left[ \left|\overline{P}_{a, y}(R,S) -  P_{a, y}(R)\right| > \frac{t}{2} \right]    +      Pr \left[\left|\frac{r}{n_{a,q}}\right|  > \frac{t}{2} \right] \\
&\le  \sum_{a \in A}\sum_{y \in Y} \left(2 \exp(\frac{- n_{a,q} t^2}{2}) + \exp(\frac{-n_{a,q} t\epsilon}{2}) \right) \\
&=  \sum_{a \in A} |Y| \left(2 \exp(\frac{- n_{a,q} t^2}{2}) + \exp(\frac{-n_{a,q} t\epsilon}{2}) \right) \\
&\le |A||Y| \left(2 \exp(\frac{- n_{min} t^2}{2}) + \exp(\frac{-n_{min} t\epsilon}{2}) \right) \\
&\le |A||Y| \left(3 \exp(\frac{- n_{min} t^2}{2}) \right) \le \delta
\end{align*}
where $n_{min}$ is the smallest $n_{a,q}$
  across all groups $S_{a,q}$.
The last step above uses the fact that  $\epsilon > t$ to simplify the term.
This fact follows from \autoref{eq:t_and_alpha}
  and uses the assumption from \autoref{thm:main_result} that $\epsilon > \frac{\alpha}{2}$.
Rearranging the term
  and then plugging in $t = \frac{\alpha}{2}$,
  we get the following lower bound for $n_{min}$:

$$\min_{a \in A} n_{a,q}  = n_{min} \ge \frac{2}{t^2} \ln{\frac{3|A||Y|}{\delta}} =  \frac{8}{\alpha^2} \ln{\frac{3|A||Y|}{\delta}}$$
\end{proof}

We next give the following upper bound on the factor
  by which number of samples increase when a privacy
  mechanism is added to conclude the proof of the theorem.
\begin{lemma}\label{eq:upper_bound}
Compared to the non-private case (\autoref{lemma:n_min_a_no_priv}),
  at most 6.34 times as many samples are needed to perform the
  audit with differential privacy guarantees (\autoref{lemma:n_min_a_priv})).
\begin{equation}
S_{dp} = \frac{\frac{8}{\alpha^2} \ln{\frac{3 |A| |Y|}{\delta}}}{\frac{2}{\alpha^2} \ln{\frac{2 |A| |Y|}{\delta}}} \le 4*\frac{\ln(3)}{\ln(2)} \approx 6.34
\end{equation}
\end{lemma}

\begin{figure}
  \centering
  \includegraphics[width=0.7\columnwidth]{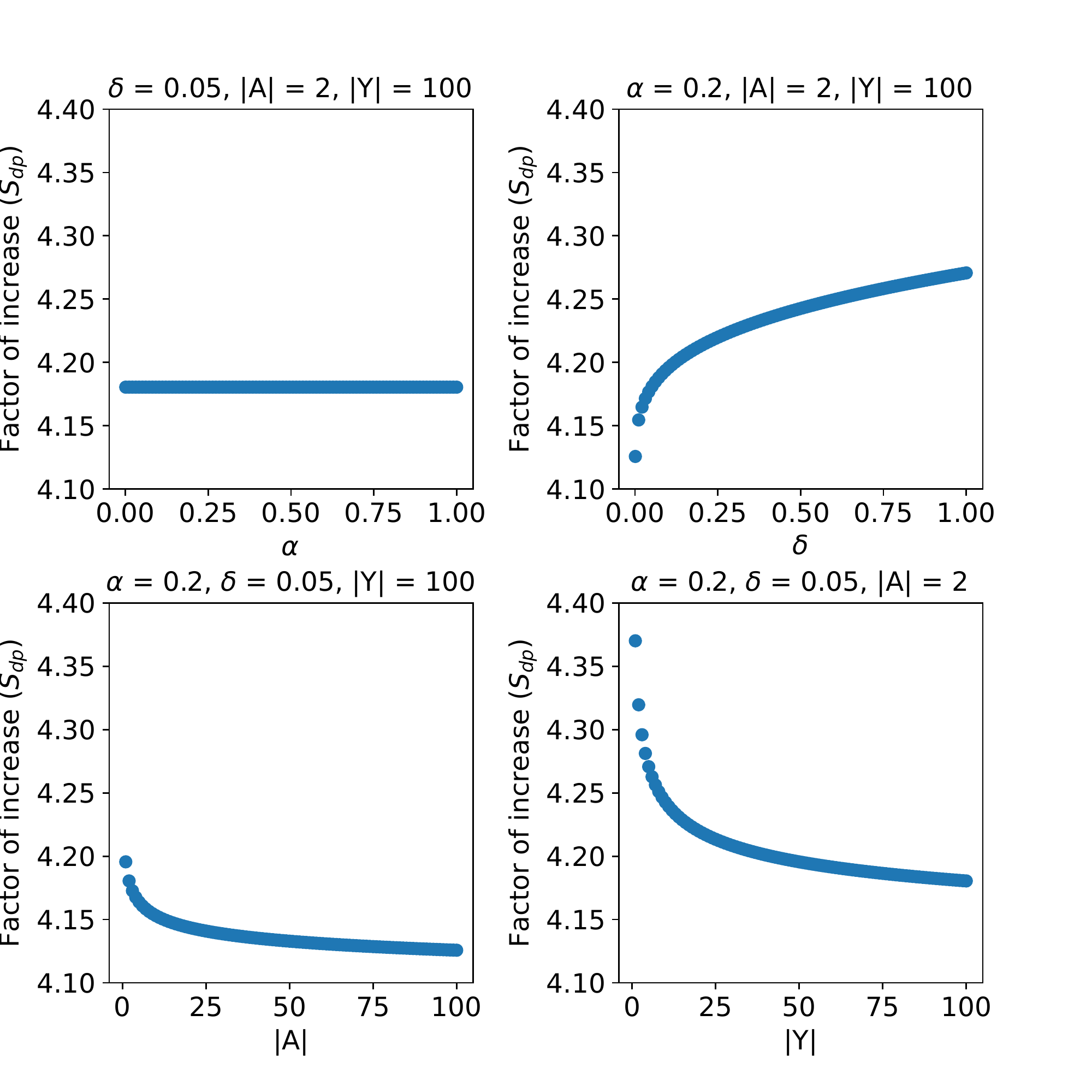}
    \caption{Relationship between auditing parameters ($\alpha$, $\delta$, $|A|$, and $|Y|$) and  the factor of increase in sample size.}
  \label{fig:sample_size_4x_bound}
\end{figure}

We prove the above lemma in~\autoref{sec:factor_upper_bound}.
While this is a strict upper bound,
  for reasonable auditing parameters,
  the overhead is much lower, around 4.
\autoref{fig:sample_size_4x_bound} shows these relationships for the
  four parameters: $\alpha$, $\delta$, $|A|$, and $|Y|$.
For all parameters, the factor of increase stays close to 4,
  lower than the true upper bound of 6.34.
We omit $\epsilon$ from the plots because the upper bound stays the same
  for any $\epsilon > \frac{\alpha}{2}$.
{ \MarkChange
\reviewfix{}
 Since a typical value for
  $\alpha$ will be close to $0$, this constraint allows for small values of $\epsilon$,
  which are known to provide reasonable privacy guarantees~\cite{dwork2014algorithmic}.
}

As an example of this more typical upper bound,
say the auditor sets the fairness gap to $\alpha=0.2$,
  a comparable parameter to the 4/5ths rule that is commonly
  applied to test for adverse impact~\cite{EEOCGuidelines}.
Assume there are 2 demographic groups ($|A| = 2$) and that relevance scores
  range from $1$ to $100$ ($|Y| = 100$), and assume the auditor would
  like to evaluate fairness with $95\%$ confidence ($\delta = 0.05$).
Then, the auditor needs a minimum of 1,879 samples from each demographic group to do the evaluation
  with privacy guarantees, compared to 450 samples without privacy,
  which is a 4.17x increase.
Such a sample size is reasonable compared to cohorts of several thousands of
  users used in prior external audits performed on
  social media platforms~\cite{Ali2019a, Imana2021},
  and is at the same order of magnitude achieved by current opt-in crowdsourcing
  efforts~\cite{CitizenBrowser}.

This small constant factor represents the increase in number of
  samples that ensuring the protection of differential privacy requires.
More importantly, it   
  demonstrates that ensuring privacy need not be a barrier
  to implementation of platform-supported auditing.
 
\section{Related Work}
  \label{sec:related_work}

As algorithmic decision making systems have become ubiquitous,
  there is a growing call for auditing them
  for potential harmful behavior.
We highlight below such work on methods for algorithmic auditing,
  their use on social media and their
  trade-offs with privacy.

\textbf{Methods for Algorithmic Auditing:}
Audits can be either internal, performed by employees
  of companies with direct access to their systems, or
  external, performed by independent third-party entities
  with usually only user-level access to the systems.
We highlight how the platform-supported auditing framework we propose
  compares to existing auditing methods.

Sandvig et al.~provides an overview and taxonomy
  of external algorithmic auditing methods~\cite{Sandvig2014}.
The taxonomy identifies five categories for types of audits:
\reviewfix{}
{\MarkChange
source code audit, non-invasive user audit, scraping audit, sock-puppet audit and crowdsourced audit.
}
Using this taxonomy,
  a recent literature review
  categorized past algorithmic audits done
  on Internet platforms~\cite{Bandy2021survey}.
Our proposal for platform-supported auditing would extend this taxonomy of audits.
It differs from source code audits because it only requires that auditors
  to have query access to algorithms' output without access to the underlying code.
It differs from the other four types of methods
  because it requires a privileged and auditor-specific query interface.

{ \MarkChange
\reviewfix{}
Some newer collaborative and user-driven auditing methods
  explored in the human-computer interaction and social computing literature
  do not directly fit into Sandvig's taxonomy.
Such methods rely on regular users to identify bias in algorithms,
  and do not formally involve auditing experts~\cite{Shen2021, DeVos2022}.
One example of newer methods is \emph{everyday algorithm auditing},
  where users of online platforms identify
  problematic behavior through their normal,
  day-to-day interactions with the platforms~\cite{Shen2021}.
Recent examples of everyday users uncovering bias in Twitter's image cropping algorithm and
  Yelp's rating system, show the ability of everyday
  users to identify problematic algorithms without
  formal or centralized auditing.
Additional work has explored making \emph{user-driven auditing}
  more effective~\cite{DeVos2022}.
Although user-driven auditing is important,
  our work instead focuses on detecting harms
  that are not visible to casual users,
  such as discriminatory ad delivery by gender or race. A user cannot know which ads they were targeted with but
  not shown due to platform's algorithmic choices based on their characteristic; thus, such studies require a broader view of the algorithm's performance.
Second, our work employs
  an auditor who can carry out
  statistical
  tests to document skew.
Such tests allow detection of more subtle differences
  that would be invisible to users.
Despite the differences,
  platform-supported auditing and user-driven audits
  can complement each other,
  with user-driven auditing suggesting the types of problems
  that may warrant more systematic, statistical evaluation 
  through our proposed framework.
}

Matias et al.~have proposed \emph{software-supported auditing},
  augmenting the effectiveness of crowdsourced audits with
  automation that chooses auditing parameters such as
  audit prompts and
  sample size~\cite{Matias2021}.
While this work rigorously estimates sample sizes,
  they do not analyze how adding a privacy guarantee changes
  the sample size required,
  something we add in \autoref{sec:privacy_tradeoffs}.
  
Reisman et al.~proposes a framework for performing
  algorithmic impact assessments and enumerates challenges
  around them~\cite{Reisman2018}.
Among other recommendations, they identify the need for
  external auditors to have meaningful
  access to periodically assess the impact of algorithms
  but they do not suggest how auditing can be done while protecting privacy.
Metaxa et al.~emphasizes the need to
  evaluate the role of personalization when auditing
  algorithmic systems~\cite{Metaxa2021}.
Our work provides a concrete proposal for how to
  implement an audit of social media platforms' personalization
  algorithms while safeguarding the privacy of users.

An audit of Pymetrics, a startup that offers a job candidate screening service, performed by external researchers in 2020 proposes
  a new \emph{cooperative audit} framework,
  where the target platform gives the auditor special access
  to its source code and data~\cite{wilson2021building}.
This framework is similar to our work in that it requires
  platform collaboration.
Our framework differs in that it requires only query access
  to the platform's algorithms, and does not require access to underlying proprietary source code and data; furthermore, it protects the privacy of the individuals participating in the audit.

\textbf{Use of Algorithm Audits on Social Media:}
Several studies have investigated the role of social
  media algorithms in biased delivery of both organic content
  and promoted ads.
Sweeney's empirical study of Google Search ads~\cite{Sweeney2013} was the first to hypothesize
  that platform-driven
  decisions can lead to discriminatory ad
  delivery; a hypothesis strengthened by evidence from subsequent works~\cite{Datta2015, gelauff2020advertising, Lambrecht2016}.
Ali and Sapiezynski et al.~confirmed this hypothesis
  by showing Facebook's algorithms skew delivery of job and housing
  ads by gender and race even when an advertiser targets a neutral audience~\cite{Ali2019a}.
  In our prior work, we
  showed how to control for job qualifications on
  Facebook and LinkedIn,
  providing evidence that skew on Facebook may be discriminatory under U.S.~law~\cite{Imana2021}.
While these studies successfully identified harms,
  each has limitations we discuss in \autoref{sec:improve_limitations}.
The new method we propose
  can be used to audit societal impacts of ad delivery algorithms
  while accounting for user privacy and other limitations.

Audits have also evaluated how social media
  algorithms bias delivery of organic content.
A sock-puppet study
  of Facebook's newsfeed, with a focus on
  content generated leading up to the Italian election in 2018,
  shows the algorithms cause ranking bias~\cite{Hargreaves2018}.
A similar sock-puppet audit compared
  reverse-chronological and algorithmic
  timelines on Twitter to show the platform's algorithms
  distort content that is shown to users~\cite{Bartley2021}.
An internal audit by Twitter also looked at the effect of
  algorithmic timelines on political content and found
  their algorithms amplify content unequally
  across the political spectrum~\cite{TwitterAlgAmplification}.
These studies quantify biases
  by comparing algorithmic and chronological timelines.
Although we do not apply our work to bias in organic content,
  our framework is generalizable to studying where
  such biases may arise from.

\textbf{Algorithmic Auditing and Privacy:}
Auditing for fairness while protecting
  privacy of users is also an active area of research
  that our work contributes to.
Segal et al.~proposed a privacy-preserving
  framework for certifying the fairness of machine learning models through
  an interactive test~\cite{Segal2021}.
Their framework protects privacy of auditors' query inputs
  by using secure computation to ensure
  the model owner does not see the data in the queries.
In contrast, our method assumes user data is
  already known to the platform,
  as is the case of social media platforms.
Our framework focuses on protecting the information
  query outputs leak about users or the platforms'
  algorithms to the auditor.

Other studies at the intersection of auditing and privacy
  have also looked at addressing
  privacy and other challenges around
  use of demographic data.
Studies by Holstein~\cite{Holstein2019} and later by Andrus~\cite{Andrus2021}
  interviewed practitioners
  from a wide range of industries to map out such challenges
  and normative questions around collection, inference, and
  use of sensitive demographics attributes of users
  for fairness efforts~\cite{Andrus2021, Holstein2019}. 
Similarly, Bogen et al.~discusses the challenges around
  access to demographic attributes that arise due
  to different laws and inconsistent practices across
  different domains such as credit, employment,
  and health~\cite{bogen2020awareness}.   
Platforms like Meta are actively working to address
  these challenges with new mechanisms
  for internal studies of the impact of sensitive attributes while protecting
  privacy~\cite{metatech, austin2021race}.
 Our proposal sidesteps these challenges as it does not require platforms to collect or
   store sensitive attributes;
   they only need to be known by the external auditor.
Similar to our work, Veale et al.'s ~proposes use of a trusted third-party entity
  to collect demographic data of users of an algorithmic system
  and later used the data for auditing the system~\cite{Veale2017}.
Our proposal differs in that it does not require collection of demographic attributes
  of all users, but just enough number needed to conduct an audit.
Other works aiming to determine disparate algorithmic outcomes based on group membership, such as by~\cite{juarez2022you} and~\cite{friedberg2022privacy}, operate under a different privacy goal -- they aim to keep group membership private from the auditor.

\section{Implications and Future work}
Privacy concerns have hindered
  increasing transparency into operation of social media platforms.
Our work addresses
  this challenge by showing it is feasible to audit relevance estimators,
  the ``brains'' of social media platforms,
  without violating the privacy of their users
  or revealing proprietary details of the platforms' algorithms.
 
  \reviewfix{}
 { \MarkChange
A natural next step for this work is
  collaboration with a social
  media platform to evaluate a prototype of our framework.
While a full implementation is future work,
   our conceptual demonstration of the framework's
   feasibility is important progress,
   suggesting proposed legislation \emph{can} be realized;
   the policy goal of external auditing with privacy is feasible.
}

Our proposal for platform-supported auditing
  gives a practical framework for implementing
  policies outlined in DSA and PATA.  
Our framework focuses on these proposals as 
  both are promising efforts to increasing
  transparency of social media platforms and their algorithms' role in influencing individuals and shaping societal discourse.
Compared to prior proposals in the U.S.~\cite{Sdaa2021, Aaa2019, Ajopta2021},
  PATA is the most comprehensive in terms of the large platforms it covers~\cite{Nonnecke2022}.
Even if PATA's ultimate fate is uncertain,
  the EU-centric DSA that has already been passed as law
  may influence future policies in the U.S. and beyond,
  similar to the way EU's GDPR has shaped the
  global privacy landscape~\cite{Linden2020}.
As an example, YouTube's announcement of the YouTube Researcher Program for researchers in more than 50 countries came on the heels of the passing of the DSA~\cite{YoutubeResearchProg}.

The scope of our framework has limitations that
   are potential avenues for future work.
 For example,
   DSA's proposal covers platforms and services other than social media
   that are outside the scope of our study.
Also, within social media platforms,
  our work focuses on how organic and promoted
  content is delivered on users' feeds,
  a place where users consume most of their content.
However, there are other features,
  such as Trends on Twitter, chosen by platforms' algorithms,
  which we do not address in our work
  but are worth studying for potential harms such as misinformation.
  
Another potential direction for future work 
  is exploring how
  platform-supported auditing can be adopted to
  study other forms of algorithmic harms.
Our example use case focuses on
  auditing for discrimination in job ad delivery.
A potential direction is exploring
  privacy mechanisms and
  metrics of fairness that
  will safeguard privacy of users
  when
  performing audits in other contexts,
  such as amplification of political and hateful content.

Our work assumes audits will be conducted under a legal framework
  that incentivizes platforms to act in good faith (\autoref{sec:model_limitations}),
  but another area of future work is to relax this assumption
  and add technical methods that look for
  accidental errors or intentional non-compliance by platforms.
Correlation of data has detected lapses in the past~\cite{FBSocialOneData}.
Technical methods, combined with the legal
  incentives proposed in DSA and PATA,
  would provide even stronger guarantees that audits are
  accurate and complete.

\raggedbottom
\section{Conclusion}
Auditing social media platforms for public interest is an
   active and pressing area of academic research, policy-making and legislation.
To address concerns raised by prior audits,
  legislations have been proposed to mandate auditing by external researchers
  without compromising privacy of platform users and business interests
  of platforms.
We propose a platform-supported auditing framework that has
  safeguards for protecting against these risks.
The center of our mechanism is \emph{increasing transparency of relevance
  estimators}, which are the core drivers of both organic and promoted content choice and prioritization
  on social media.
Our analysis shows privacy-preserving auditing of relevance estimators
  can be implemented with high statistical confidence,
  provided that the sample size is increased by a small constant factor.
Our findings offer a novel technical solution for how to practically
  implement public oversight of social media companies,
  a core goal the proposed legislations are pushing for.
  
\section{Acknowledgments}
We thank anonymous reviewers for their constructive feedback. This work was funded in part by a grant from The Rose Foundation and NSF awards \#1916153, \#1956435, and \#1943584. 

%

\bibliographystyle{acm}

\received{July 2022}
\received[revised]{October 2022}
\received[accepted]{January 2023}

%
\appendix

\section{Number of Samples Required for Auditing Without Privacy}
  \label{sec:non_private_proof}

In~\autoref{sec:main_result}, we suggest~\autoref{lemma:n_min_a_no_priv} holds
  as lower bound for number of samples required for auditing without privacy.
Here we give a detailed proof for~\autoref{lemma:n_min_a_no_priv}.
Our proof follows Segal et al.'s work~\cite{Segal2021},
  with modifications to adopt it to our
  use-case that considers qualification as an additional attribute of users.

\begin{proof}
An auditor uses a sample set $S$ of users to perform an audit.
Consider the following ``bad'' event where the sampling error is above some threshold $t > 0$:
$$\overline{P}_{a, y}(R,S)\textrm{ is bad if}:~ |\overline{P}_{a, y}(R,S) -  P_{a, y}(R)| > t.$$

We would like to bound the probability of this event for all demographic groups in $A$ and possible outputs in $Y$:
$$Pr [\exists a\in A ~\text{and}~ y \in Y: \overline{P}_{a, y}(R,S)~\text{is bad}] \le \delta$$

We use union bound followed by Hoeffding's concentration bound:
\begin{align*}
& Pr [\exists a\in A ~\text{and}~ y \in Y: \overline{P}_{a, y}(R,S)~\text{is bad}] \\
&\le \sum_{a \in A}\sum_{y \in Y} Pr \left[ \overline{P}_{a, y}(R,S)~\text{is bad} \right]  \\
&= \sum_{a \in A}\sum_{y \in Y} Pr \left[ |\overline{P}_{a, y}(R,S) -  P_{a, y}(R)| > t \right] \\
&\le  \sum_{a \in A}\sum_{y \in Y} 2 \exp(-2 n_{a,q} t^2) \\
&= \sum_{a \in A} |Y| 2 \exp(-2 n_{a,q} t^2) \\
&\le |A||Y| 2 \exp(-2 n_{min} t^2)
\end{align*}

where $n_{min}$ is the number of people in a group in $S$
  that has least number of qualified people.
We want the above probability to be small, i.e., $|A||Y| 2 \exp(-2 n_{min} t^2) \le \delta$.
Rearranging, we get the following bound on $n_{min}$:

\begin{equation}\label{eq:n_min_t}
n_{min} \ge \frac{1}{2t^2} \ln{\frac{2|A||Y|}{\delta}}
\end{equation}

We next derive the value of $t$ needed to guarantee $(\alpha, \delta)$-fairness.
Based on Definition  \autoref{def:alpha_fair},
  it is sufficient to show that,
  for any pair $a_1, a_2 \in A$ and any $y \in Y$,
  the fairness gap is bounded by $\alpha$:

$$ |P_{a_1, y}(R) - P_{a_2, y}(R)| \le \alpha $$

Conditioning on the above bad event not occurring,
  we start with $|P_{a_1, y}(R) - P_{a_2, y}(R)|$ and
  apply triangle inequality.
In the second inequality below,
  we add the term $( EFG(R, S) - |\overline{P}_{a_1, y}(R, S) - \overline{P}_{a_2, y}(R, S)| )$
  because it is positive (based on definition of EFG).

\begin{align*}
& |P_{a_1, y}(R) - P_{a_2, y}(R)| \\
&\le |P_{a_1, y}(R)|  +  |P_{a_2, y}(R)| \\
& \le |P_{a_1, y}(R)|  +  |P_{a_2, y}(R)| + ( EFG(R, S) - |\overline{P}_{a_1, y}(R, S) - \overline{P}_{a_2, y}(R, S)| ) \\
& \le |P_{a_1, y}(R)|  +  |P_{a_2, y}(R)| +  EFG(R, S) - ( |\overline{P}_{a_1, y}(R, S)| + |\overline{P}_{a_2, y}(R, S)|  ) \\
& = |P_{a_1, y}(R)|- |\overline{P}_{a_1, y}(R, S)|  +  |P_{a_2, y}(R)| - |\overline{P}_{a_2, y}(R, S)| +  EFG(R, S) \\
& \le t + t + EFG(R, S)
\end{align*}

We want $2t + EFG(R, S) \le \alpha$. Therefore, $t \le \frac{\alpha - EFG(R,S)}{2}$.
For $EFG(R,S) \le \alpha$, $t \le \frac{\alpha - EFG(R,S)}{2} \le \frac{\alpha}{2}$ holds.
Plugging in the value of $t = \frac{\alpha}{2}$ to \autoref{eq:n_min_t}
  gives us a lower bound for the number of samples
  needed:
$$n_{min} \ge \frac{2}{\alpha^2} \ln{\frac{2|A||Y|}{\delta}} $$

\end{proof}

\section{Upper bound on increase of number of samples}
  \label{sec:factor_upper_bound}

In this appendix  
  we give a proof for~\autoref{eq:upper_bound} to provide
  an upper bound on the factor
 by which number of samples increase when a privacy
 mechanism is added.
 
 \begin{proof}
Let $P = \ln{\frac{|A||Y|}{\delta}}$. Then:

\begin{align*}
 S_{dp} = \frac{\frac{8}{\alpha^2} \ln{\frac{3 |A| |Y|}{\delta}}}{\frac{2}{\alpha^2} \ln{\frac{2 |A| |Y|}{\delta}}} = 4 \left(\frac{\ln{\frac{3 |A| |Y|}{\delta}}}{\ln{\frac{2 |A| |Y|}{\delta}}} \right) = 4 \left(\frac{\ln{3} + \ln{\frac{|A| |Y|}{\delta}}}{\ln{2} + \ln{\frac{|A| |Y|}{\delta}}} \right)
= 4 \left(\frac{\ln{3} + P}{\ln{2} +P} \right)
\end{align*}

Because $\delta$ is a probability and $A$ and $Y$ cannot
  be empty, we know $|A| \ge 1$, $|Y| \ge 1$, and $\delta \le 1$.
Therefore, it is always the case that $\frac{|A||Y|}{\delta} \ge 1$ and $P \ge 0$.

Now, consider $f(P) = 4 \left(\frac{\ln{3} + P}{\ln{2} +P} \right)$.
Because $P \ge 0$, $f(P)$ is maximized when $P = 0$, and monotonically decreases as
  $P$ increases. Therefore, $f(P) \le f(0)$ for all $P \ge 0$. Finally:

\begin{align*}
 \frac{\frac{8}{\alpha^2} \ln{\frac{3 |A| |Y|}{\delta}}}{\frac{2}{\alpha^2} \ln{\frac{2 |A| |Y|}{\delta}}} = 4 \left(\frac{\ln{3} + P}{\ln{2} +P} \right) = f(P) \le f(0) =  4\times\frac{\ln(3)}{\ln(2)} \approx 6.34
\end{align*}
\end{proof}

\section{Applying Hoeffding's}
  \label{sec:hoef_with_bias}

In~\autoref{sec:main_result}, we use Hoeffding's inequality to bound~\autoref{eq:two_error_terms}.
Here we give a proof for why we can apply Hoeffding's inequality even in the presence of potential bias in $R$.

From~\autoref{eq:two_error_terms}, we would to apply Hoeffding's bound to the following sampling error term: 
$$Pr \left[ \left|\overline{P}_{a, y}(R,S) -  P_{a, y}(R)\right| > \frac{t}{2} \right]$$
Hoeffding's inequality gives
  an upper bound on the probability that the sum of bounded
  random variables deviates from its expected value~\cite{Hoeffding63a}.

To apply Hoeffding's, we need to show sampling is i.i.d.~and
  that we are summing bounded random variables.
An auditor can sample i.i.d.~in
  several ways: the platform may provide sampling or
  the auditor may use an external source of a unique set of users.
Based on~\autoref{eq:p_a_y}, $\overline{P}_{a, y}(R,S)$ is a sum of $n_{a,q}$ indicator
  variables defined on each sample in $S_{a,q}$.
Indicator variables can only
  hold a value of $0$ or $1$, so they are bounded.
The remaining requirement we need to show to apply Hoeffding's is:
\begin{equation}\label{eq:hoeff_req}
E[\overline{P}_{a, y}(R,S)] = P_{a, y}(R)
\end{equation}

The goal of the auditor is to test
  for potential bias that is correlated with some sensitive attribute.
We next show $\overline{P}_{a, y}(R,S)$ is unbiased estimator of $P_{a, y}(R)$
  (by showing~\autoref{eq:hoeff_req} holds) even in the presence of bias per group as long
  as the samples in $S_{a,q}$ are i.i.d\@.
 We consider three cases: when bias is
   an additive constant factor,
   a multiplicative constant factor, and
   an additive discrete random variable.
  
\paragraph{Bias that is an additive, constant factor}
Consider the following formulation that takes such bias into account:
\begin{equation}\label{eq:R_with_bias}
R(x) = T(x) + b_a
\end{equation}
where $b_a$ is a constant bias
  for a user with attribute $a$, and
  $T(x)$ is a random variable reflecting that individual $x$'s history.

As mentioned before, $S_{a,q}$ represent subset of $S$ with given values of $a$ and $q$.
We consider the subset of qualified individuals so $q=1$.
Let  $\overline{S_{a,q}}$ represent the complement of $S_{a,q}$.

\begin{align}
 E[\overline{P}_{a, y}(R,S)] \nonumber &= E\left[ \frac{1}{n_{a,q}} \sum_{i=1}^{|S|} \mathbb{1}\{ R(x_i) = y \land a_i = a  \land q_i = 1 \} \right] \nonumber \\
&= E\left[ \frac{1}{n_{a,q}} \sum_{i=1}^{|S_{a,q}|} \mathbb{1}\{ R(x_i) = y  \}  + \sum_{i=1}^{|\overline{S_{a,q}}|} 0  \right] \nonumber \text{\quad.....separate $S_{a,q}$ and $\overline{S_{a,q}}$ } \\ 
&= \frac{1}{n_{a,q}} \sum_{i=1}^{|S_{a,q}|} E\left[ \mathbb{1}\{ R(x_i) = y \} \right] \nonumber \\ 
&= \frac{1}{n_{a,q}} \sum_{i=1}^{|S_{a,q}|} 0 * Pr[R(x_i) \neq y ] + 1 * Pr[R(x_i) = y ]  \nonumber \\
&= \frac{1}{n_{a,q}} \sum_{i=1}^{|S_{a,q}|}  Pr[R(x_i) = y] \label{eq:r_x_i_y} \\
&= \frac{1}{n_{a,q}} \sum_{i=1}^{|S_{a,q}|}  Pr[T(x_i) + b_a = y ]  \nonumber \\ 
&= \frac{1}{n_{a,q}} \sum_{i=1}^{|S_{a,q}|}  Pr[T(x_i) = y - b_a ]  \nonumber \\ 
&= \frac{1}{n_{a,q}} \sum_{i=1}^{|S_{a,q}|} {P}_{a, y-b_a}(T) \nonumber  \text{\qquad \qquad \qquad \qquad.....by i.i.d. assumption} \\
&= \frac{n_{a,q}}{n_{a,q}} {P}_{a, y-b_a}(T) \nonumber \\
&= {P}_{a, y-b_a}(T) \nonumber \\
&= {P}_{a, y}(R) \nonumber   \text{\qquad\qquad\qquad\qquad.....plug in~\autoref{eq:R_with_bias} in~\autoref{eq:true_P_a_r}} \\\nonumber
\end{align}  

Therefore, we can apply Hoeffding's for samples in a group even if the group
  attribute induces an additive bias.

\paragraph{Bias that is a multiplicative, constant factor}
One can follow similar steps to show~\autoref{eq:hoeff_req} holds for the case a multiplicative constant bias.
Let
\begin{equation}\label{eq:R_with_bias_mul}
R(x) = T(x) * b_a
\end{equation}
where $b_a$ is a constant.

\begin{align*}
E[\overline{P}_{a, y}(R,S)]  &= \frac{1}{n_{a,q}} \sum_{i=1}^{|S_{a,q}|}  Pr[R(x_i) = y] \qquad \text{.......by~\autoref{eq:r_x_i_y}}  \\
&= \frac{1}{n_{a,q}} \sum_{i=1}^{|S_{a,q}|}  Pr[T(x_i) * b_a = y]  \\ 
&= \frac{1}{n_{a,q}} \sum_{i=1}^{|S_{a,q}|}  Pr[T(x_i) = \frac{y}{b_a}]  \\ 
&= \frac{1}{n_{a,q}} \sum_{i=1}^{|S_{a,q}|} {P}_{a, \frac{y}{b_a}}(T) \qquad \text{\quad.......by i.i.d. assumption} \\
&= \frac{n_{a,q}}{n_{a,q}} {P}_{a, \frac{y}{b_a}}(T) \\
&= {P}_{a, \frac{y}{b_a}}(T) \\
&= {P}_{a, y}(R) \text{\qquad.....plug in~\autoref{eq:R_with_bias_mul} in~\autoref{eq:true_P_a_r}} \\
\end{align*}

\paragraph{Bias that is a random variable (not a constant)}
Consider bias that is a discrete random variable and is an additive factor.
Let $R(x) = T(x) + B_a$ where $B_a$ is a discrete random variable.
We would like to show~\autoref{eq:hoeff_req} holds for this case.
We look at each side of the equation separately:
\begin{align} 
E[\overline{P}_{a, y}(R,S)] \nonumber &= \frac{1}{n_{a,q}} \sum_{i=1}^{|S_{a,q}|}  Pr[R(x_i) = y]  \qquad \text{\qquad\qquad\qquad.......by~\autoref{eq:r_x_i_y}} \nonumber \\
&= \frac{1}{n_{a,q}} \sum_{i=1}^{|S_{a,q}|}  Pr[T(x_i) + B_a = y]  \nonumber \\
&= \frac{1}{n_{a,q}} \sum_{i=1}^{|S_{a,q}|} \sum_{b}  Pr[T(x_i) = y - b ] * Pr[B_a = b]  \nonumber\\
&= \frac{1}{n_{a,q}}  \sum_{b} \sum_{i=1}^{|S_{a,q}|}  Pr[T(x_i) = y - b] * Pr[B_a = b]  \nonumber\\
&= \frac{1}{n_{a,q}}  \sum_{b} Pr[B_a = b]  \sum_{i=1}^{|S_{a,q}|}  Pr[T(x_i) = y - b]  \nonumber\\
&= \frac{1}{n_{a,q}}  \sum_{b} Pr[B_a = b]  \sum_{i=1}^{|S_{a,q}|}  P_{a,y-b}(T) \qquad \text{.......by i.i.d. assumption} \nonumber\\
&= \frac{1}{n_{a,q}}  \sum_{b} Pr[B_a = b]  * n_{a,q}  * P_{a,y-b}(T) \nonumber\\
&= \sum_{b} Pr[B_a = b] * P_{a,y-b}(T) \label{eq:lhs} \\\nonumber
\end{align} 

\begin{align}
{P}_{a, y}(R) \nonumber &=  Pr_{(x, a', q) \sim\ X} \left[  R(x) = y | a' = a  \land q = 1  \right] \nonumber \\
&=  Pr_{(x, a', q) \sim\ X} \left[  T(x) + B_a = y | a' = a  \land q = 1  \right] \nonumber \\
&=  \sum_{b} Pr[B_a = b] * Pr_{(x, a', q) \sim\ X} \left[  T(x)  = y - b | a' = a  \land q = 1  \right] \nonumber\\
&=  \sum_{b} Pr[B_a = b] * P_{a, y-b}(T)  \label{eq:rhs} \\\nonumber
\end{align}  

Since~\autoref{eq:lhs} and~\autoref{eq:rhs} are equal, $E[\overline{P}_{a, y}(R,S)] = {P}_{a, y}(R) $.
  
\end{document}